\documentclass[12pt]{article}
\usepackage{amssymb,amsmath}
\usepackage{graphicx}
\usepackage{ccaption}
\usepackage[usenames]{color}
\usepackage{mathrsfs}

\pdfoutput=1

\usepackage[
      colorlinks=true,
      linkcolor=blue,
      urlcolor=blue,
      filecolor=blue,
      citecolor=red,
      pdfstartview=FitV,
      pdftitle={},
        pdfauthor={Donald Marolf, Mukund Rangamani, Mark Van Raamsdonk},
        pdfsubject={},
        pdfkeywords={},
        pdfpagemode=None,
        bookmarksopen=true
      ]{hyperref}

\makeatletter
\@addtoreset{equation}{section}

\makeatletter
\renewcommand\section{\@startsection {section}{1}{\z@}%
                                   {-3.5ex \@plus -1ex \@minus -.2ex}
                                   {2.3ex \@plus.2ex}%
                                   {\normalfont\large\bfseries}}
\renewcommand\subsection{\@startsection{subsection}{2}{\z@}%
                                     {-3.25ex\@plus -1ex \@minus -.2ex}%
                                     {1.5ex \@plus .2ex}%
                                     {\normalfont\bfseries}}

  \captionnamefont{\bfseries}
  \captiontitlefont{\small\sffamily}
  \captiondelim{: }
  \hangcaption

\def\sec#1{\S\ref{#1}}
\def\fig#1{Fig.\,\ref{#1}}
\def\req#1{(\ref{#1})}
\def\App#1{Appendix \ref{#1}}

\def\dS#1{dS$_{#1}$}
\def\AdS#1{AdS$_{#1}$}
\def\ESU#1{\text{ESU}$_{#1}$}
\def\Sp{{\bf S}}

\newcommand{\nbox}{{\,\lower0.9pt\vbox{\hrule \hbox{\vrule height 0.2 cm \hskip 0.19 cm \vrule height 0.2 cm}\hrule}\,}}
\newcommand{\Tr}{\ {\rm Tr}\ }

\definecolor{purple}{rgb}{0.8,0.3,0.5}
\definecolor{green}{rgb}{0.1,0.8,0.2}

\title{{\bf \Large Holographic models of de Sitter QFTs}}
\author{\normalsize
Donald Marolf$^{\,a}$\footnote{marolf@physics.ucsb.edu},
Mukund Rangamani$^b$\footnote{mukund.rangamani@durham.ac.uk}, Mark Van Raamsdonk$^c$\footnote{mav@phas.ubc.ca} \\ \\
$^a$ \small \sl Physics Department, UCSB, Santa Barbara, CA 93106, USA. \\
\small \sl $^{b}$  Centre for Particle Theory \& Department of
Mathematical Sciences,
\\[-1.5mm]
\small \sl Science Laboratories, South Road, Durham DH1 3LE, United Kingdom. \\
\small{\emph{$^{c}$   Department of
Physics and Astronomy, University of British Columbia}}, \\[-1.5mm]
\small{\emph{6224 Agricultural Road, Vancouver, B.C., V6T 1W9, Canada}}
}

\begin{document}

\setlength{\baselineskip}{16pt}
\begin{titlepage}
\maketitle
\begin{picture}(0,0)(0,0)
\put(380, 300){DCPT-10/27}
\end{picture}
\vspace{-36pt}

\begin{abstract}
We describe the dynamics of strongly coupled field theories in de Sitter spacetime using the holographic gauge/gravity duality. The main motivation for this is to explore the possibility of dynamical phase transitions during cosmological evolution. Specifically, we study two classes of theories: (i) conformal field theories on de Sitter in the static patch which are maintained in equilibrium at temperatures that may differ from the de Sitter temperature and (ii) confining gauge theories on de Sitter spacetime. In the former case we show the such states make sense from the holographic viewpoint in that they have regular bulk gravity solutions. In the latter situation we add to the evidence for a confinement/deconfinement transition for a large $N$ planar gauge theory as the cosmological acceleration is increased past a critical value. For the field theories we study, the critical acceleration corresponds to a de Sitter temperature which is {\it less} than the Minkowski space deconfinement transition temperature by a factor of the spacetime dimension.
 \end{abstract}
\thispagestyle{empty}
\setcounter{page}{0}
\end{titlepage}

\renewcommand{\thefootnote}{\arabic{footnote}}
\tableofcontents

\section{Introduction and Summary}
\label{s:intro}

Quantum field theories on curved spacetimes are of immense physical importance in many circumstances, for example in deriving the predictions of inflationary cosmology or in understanding the semi-classical physics of black holes. Historically, much of our understanding of curved-spacetime quantum field theory has been based on a study of free fields. Already in this context one encounters interesting physical issues such as the definition of the vacuum state and particle production due to spacetime curvature, cf., \cite{Birrell:1982ix}. While  there has been some work extending this to perturbative studies of weakly coupled field theories, direct studies of strongly coupled field theories pose technical and conceptual challenges.

In the context of cosmological evolution, one could readily envision the importance of understanding the dynamics of strongly interacting field theories on cosmological backgrounds. In fact, as we now explain, there is good reason to believe that qualitatively new phenomena arise in such theories, and that some of these phenomena might indeed be relevant for the physics of the early universe.

As an example, consider the case of de Sitter spacetime, characterized by a Hubble parameter $H$ which parameterizes the rate of  accelerated expansion. From studies of free field theory in de Sitter space, it is well known that an observer moving along a timelike geodesic observes a thermal bath of particles at temperature $T_\text{dS} = \frac{1}{2\pi}\, H $ if the corresponding (free) field is in its de Sitter-invariant vacuum state \cite{Gibbons:1977mu}. This thermal behaviour is intimately associated with the cosmological horizon seen by the observer. While this resembles the thermal character of black hole horizons, there is a crucial difference: the cosmological horizon is observer-dependent. Nevertheless, we can associate a temperature to de Sitter spacetime and one should therefore consider local quantum theory residing on this background to be in an appropriate density matrix inside any horizon.

A natural question that we can then ask is the following: consider interacting field theories which (in Minkowski space) exhibit multiple phases at non-zero temperatures, and undergo phase transitions as their temperature is increased or decreased past some critical value $T_c$. When such theories are coupled to a background de Sitter metric, can a change in the cosmological scale - and thus a change in $T_{dS}$ - produce a similar phase transition? Specifically, one might expect to find a phase transition as we vary $H$ past a critical value where $T_\text{dS} = T_c$.

We should emphasize that while plausible, the existence of such phase transitions is by no means obvious. In the de Sitter background, the temperature $T_\text{dS}$ and the curvature scale $R_\text{dS}$ (which is inversely proportional to the Hubble constant $H$) are linked by $T_\text{dS} \sim 1/R_\text{dS}$, which implies that one cannot easily disentangle the effects of curvature from thermal effects.  Indeed, the state of the field theory in a curved background like de Sitter could be quite different from the state at the same nominal temperature in Minkowski space (where we have the best developed intuition for various phases). It is therefore a nontrivial question whether in the context of inflationary cosmology ``thermal'' phase transitions can occur dynamically as the acceleration of the universe changes.

In this paper, our approach will be to study this question using holographic gauge/gravity duality \cite{Maldacena:1997re}. Specifically, we consider field theories which may be studied using a gravity dual and which exhibit a confined phase at low temperatures and a deconfined phase at high temperatures. Using holographic methods, we can ask whether these theories also exhibit deconfinement as the acceleration is increased on a de Sitter background. The confining gauge theories we are interested in are defined by Scherk-Schwarz compactification of CFTs on a circle, so the study of these theories in de Sitter spacetime is equivalent to a holographic study of CFTs on \dS{d} $ \times\, \Sp^1_{SS}$.

Fortunately, this system has been considered in detail before, with the relevant dual gravity solutions derived and discussed (including some comments about possible field theory interpretation) in \cite{Aharony:2002cx, Balasubramanian:2002am, Cai:2002mr, Ross:2004cb, Balasubramanian:2005bg, He:2007ji, Hutasoit:2009xy}. In particular, the work \cite{Hutasoit:2009xy} emphasized that such theories indeed exhibit phase transitions as the ratio between de Sitter radius and $\Sp^1_{SS}$ radius is varied, and that small $R_{\Sp^1}/R_{\text{dS}}$ and large $R_{\Sp^1}/R_{\text{dS}}$ phases exhibit properties of a hadronized phase and a plasma phase, respectively.\footnote{Towards the end of the introduction we review in some detail earlier works that utilize the gauge/gravity correspondence to learn about strongly coupled theories in de Sitter spacetime.} Interpreting these results in terms of our confining gauge theory (i.e. the CFT on a fixed-size $\Sp^1_{SS}$), they confirm our expectations that there is a phase transition at a critical value of the de Sitter acceleration, and strongly suggest that this phase transition is associated with deconfinement. On the other hand, we show that the transition does not occur when the de Sitter temperature $T_\text{dS}$ equals the Minkowski space transition temperature $T_c$, but rather at a lower temperature $T_\text{dS} = T_c/d$.

In order to relate the the de Sitter transition to the Minkowski space deconfinement transition more directly, we note that it is useful to consider varying the field theory temperature $T$ and the Hubble parameter $H = 2\pi \,T_\text{dS}$ independently. Previous work focused on field theory states defined on global de Sitter space, implying the restriction $T = T_\text{dS}$. However, by working in the static patch of de Sitter, we will be able to decouple the field theory temperature from the Hubble constant and thus to provide evidence that the large $T = T_\text{dS}$ phase is indeed continuously connected to the deconfined plasma in Minkowski space in the $T$-$H$ phase diagram (see figure \fig{f:confdeconf}).

Since $T/H$ provides a dimensionless parameter that is physically relevant even for conformal field theories, it is also interesting to study conformal theories on static patch de Sitter as a function of this ratio. For CFTs with gravity duals, we are able to perform this analysis quite explicitly, though we do not find any phase transitions as $T/H$ is varied.

\paragraph{Methodology and Outline:}

We remind the reader that gauge/gravity duality relates dynamics of strongly coupled (non-gravitational) conformal field theories to the dynamics of string theory or gravity on the class of asymptotically locally AdS spacetimes. Typically, this has been employed to study strongly coupled conformal field theories on Minkowski space or on the Einstein Static Universe (ESU). However, it is also possible to use the duality to study certain non-conformal field theories on more general spacetimes, including cosmological backgrounds such as de Sitter spacetime.

The key ingredient we will use is that when we consider a conformal field theory on a non-dynamical background spacetime ${\cal B}_d$, the gauge/gravity correspondence maps the problem in a suitable limit to dynamics of pure gravity in an asymptotically locally AdS spacetime ${\cal M}_{d+1}$. This bulk spacetime manifold has as its conformal boundary ${\cal B}_{d}$ on which we will pick our chosen metric $\gamma_{\mu\nu}$. The states of the CFT are then dual to bulk geometries ${\cal M}_{d+1}$ with these boundary conditions.  When more than one such spacetime represents an equilibrium state at a given temperature $T$, the CFT has multiple phases. This set-up has recently been exploited to analyze the behaviour of strongly coupled field theories in black hole backgrounds \cite{Hubeny:2009ru,Hubeny:2009kz,Hubeny:2009rc}. In these papers the aim was to understand the key features of Hawking radiation of strongly coupled quanta. The problem at hand  is simpler since we wish to understand the dynamics in de Sitter spacetime which has more symmetry than a black hole background, and at least in some cases one is able to find the dual gravity solutions of interest quite explicitly.

To facilitate the discussion we will concentrate on two distinct classes of theories. We first describe the behaviour of strongly coupled conformal field theories on dS$_d$ which are somewhat simpler to analyze. Subsequently, we turn to confining theories on \dS{d}; the field theories in question are obtained via Scherk-Schwarz compactification of higher dimensional field theories and we describe how they might undergo a confinement-deconfinement phase transition as we vary the background cosmological constant.

After a brief review of some relevant properties of de Sitter spacetime in \sec{s:bkgnd}, we warm up in \sec{s:cftds}  by using AdS/CFT techniques to study the physics of conformal field theories on de Sitter spacetime. For the de Sitter-invariant vacuum state in a CFT (whose correlation functions are obtained by analytic continuation from correlation function on the sphere) there can be no interesting dependence on the Hubble parameter, since $H$ provides the only scale in the problem. Therefore one expects no interesting phase transitions when we consider CFT on global dS spacetime.

However, as we will explain in detail,  within the static patch of dS spacetime we can study field theories in thermal states corresponding to temperatures  $T$ different than the de Sitter temperature $T_\text{dS}$. These states will be singular on the global dS spacetime, but that singularity will be hidden behind the cosmological horizon of the static patch. Indeed while the existence of such states is clear for free field theories, we will show that they in fact also make sense for strongly coupled CFTs. In particular, we find a one-parameter family of regular dual gravity solutions corresponding to conformal field theories at arbitrary values of the dimensionless combination $T/H$. These interpolate between the solution dual to the CFT at finite temperature on Minkowski space and the solution dual to the dS spacetime CFT in the de Sitter vacuum state. Using the dual gravity solutions, we calculate the stress-energy tensor in the field theory for general $T$ and $H$. This stress tensor will be divergent on the cosmological horizon for all values of $T \neq T_\text{dS}$ but will be manifestly regular when the temperature of the static patch coincides with the dS temperature.

The gravitational dual solutions of these field theories are interesting in their own right - they are completely smooth and have a regular Killing horizon with constant surface gravity. The surface gravity is in fact related to the temperature of the field theory $T$. At the same time this bulk Killing horizon asymptotically approaches the boundary de Sitter horizon (of the static patch), which has associated temperature $T_\text{dS}$. While classical gravity solutions with multiple Killing horizons at different temperatures are common (the Schwarzschild de Sitter spacetime being a prime example), we have here a unique situation where the bulk spacetime horizon has a different temperature\footnote{In fact similar solutions can be constructed for black hole geometries on the boundary. For instance there is a one-parameter extension of the three dimensional black funnel in \AdS{3} (with the two dimensional black hole as the boundary) described in \cite{Hubeny:2009ru} which have a bulk temperature different from the boundary black hole's Hawking temperature.} from the horizon in the boundary despite the fact that the two are continuously connected.

Having gained intuition from our analysis of CFTs, we turn in \sec{s:confdS} to the more interesting case of confining gauge theory on de Sitter space. While one would ideally like to consider a confining gauge theory like QCD on dS for physical interest, one is stymied by the absence of a holographic dual for QCD. To obtain a confining gauge theory that can be studied using holographic methods, we consider a conformal field theory compactified on a circle with anti-periodic boundary conditions for fermions (i.e. Scherk-Schwarz compactification). As we will review, the resulting lower-dimensional theory is confining, and in particular, undergoes a deconfinement phase transition at a temperature of order the Kaluza-Klein scale. Thus, to study confining gauge theory on \dS{d-1}, one need only study conformal field theory on $\Sp^1_{SS} \times$ \dS{d-1}. The scale of confinement is simply the radius $R$ of the $\Sp^1_{SS}$, and we will keep this  fixed as we vary the Hubble parameter of the de Sitter space. The equilibrium state of the field theory for a given $H$ is the state dual to the asymptotically locally \AdS{d+1} geometry  whose boundary geometry is $\Sp^1_{SS} \times$ \dS{d-1}.  As noted in \cite{Hutasoit:2009xy} following earlier work \cite{Aharony:2002cx, Balasubramanian:2002am, Cai:2002mr, Ross:2004cb, Balasubramanian:2005bg, He:2007ji}, it is easy to see using the Euclidean picture that one indeed has a non-trivial phase transition in this set-up. Since the Euclidean continuation is an AdS spacetime with boundary $\Sp^{d-1} \times \Sp^1_{SS}$, this is nothing but the familiar Hawking-Page transition \cite{Hawking:1982dh,Witten:1998zw}. The current set-up is simply a double Wick rotation of the situation discussed in \cite{Witten:1998zw}; one starts with $\Sp^{d-1} \times \Sp^1_{SS}$ and analytically continues the $\Sp^{d-1}$ into \dS{d-1} in contrast to the usual Hawking-Page story where one analytically continues the $\Sp^1_{SS}$ to obtain a Lorentzian spacetime.

 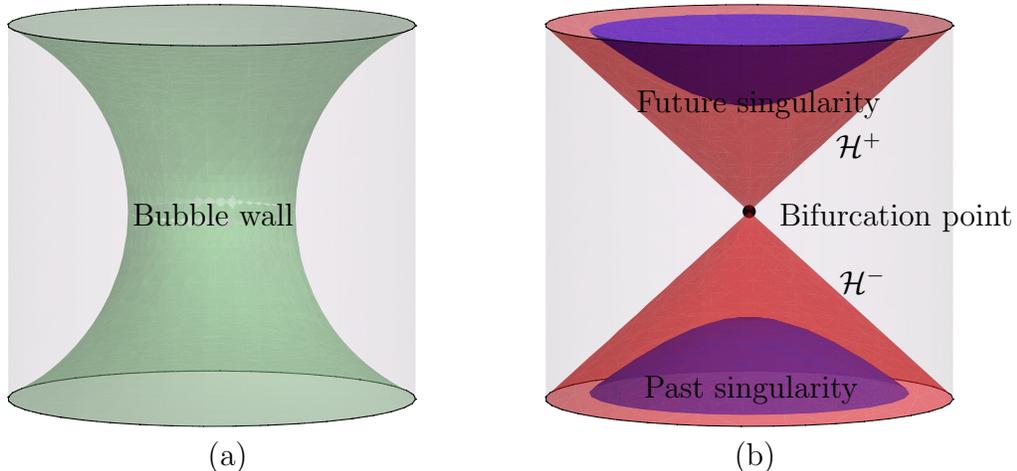
\begin{figure}[t]
\begin{center}
\includegraphics[scale = 0.65]{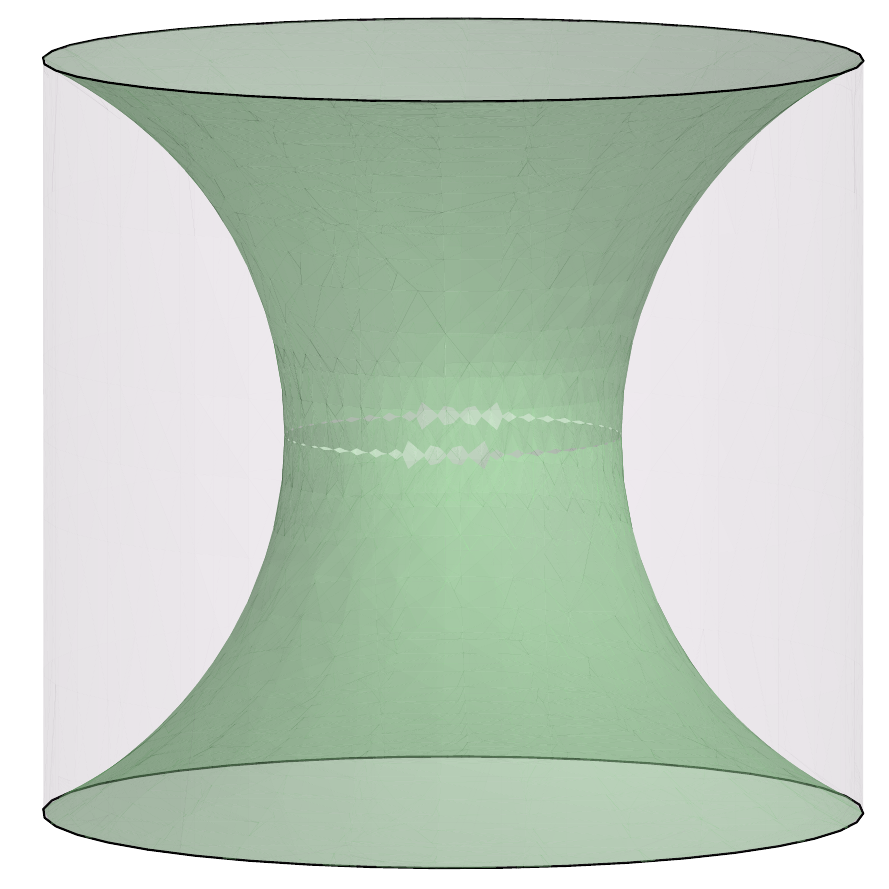} \hspace{1cm}
\includegraphics[scale= 0.65]{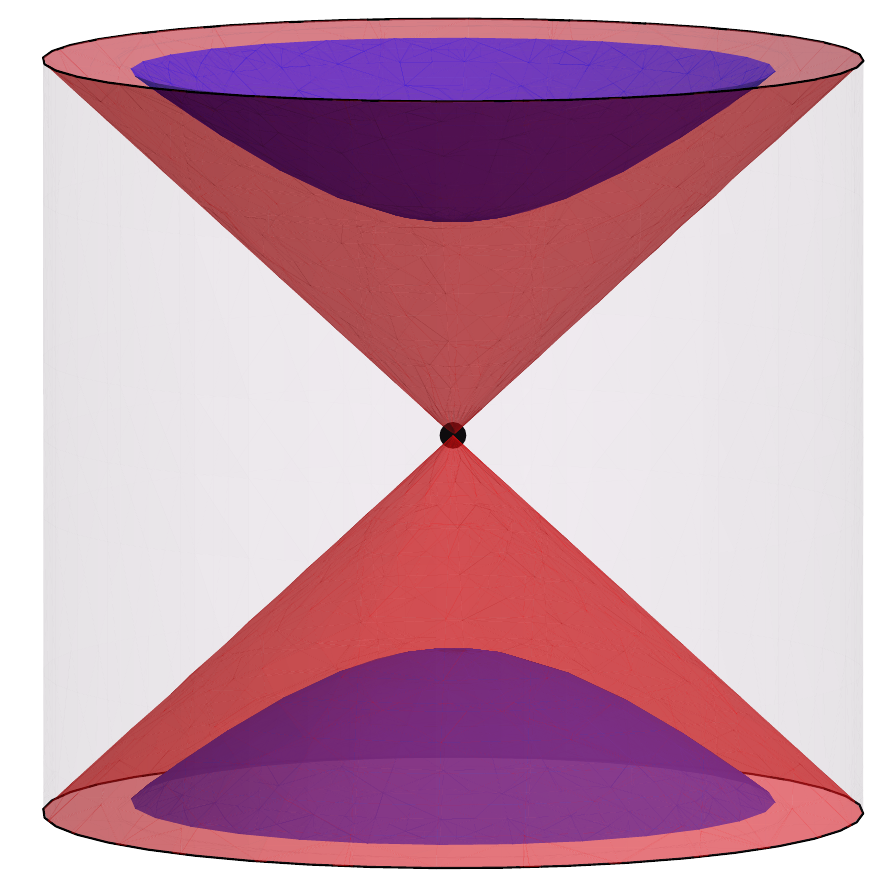}
 \caption{ The geometries relevant for holographic descriptions of confining theories on de Sitter spacetime: (a) the bubble of nothing spacetime in AdS and (b) the Ba\~nados black hole.
 The outer cylinder in both cases is the AdS boundary. For the bubble of nothing, the part of the spacetime behind the bubble whose trajectory is shown is not part of the spacetime manifold.
 For the Ba\~nados black hole we have sketched the causal diagram: the singularities are hyperbolae located behind an event horizon ${\cal H}^+ \cup {\cal H}^-$. The curious feature of this solution is that it has a bifurcation point, as opposed to a regular bifurcation surface encountered for more familiar black hole spacetimes.
 .}
 \begin{picture}(0,0)
\setlength{\unitlength}{1cm}
\put(-4.5,7.5){Bubble wall}
\put(4.1,7.5){Bifurcation point}
\put(2.2,9.0){Future singularity}
\put(2.3, 5.2){Past singularity}
\put(4.85,8.4){${\cal H}^+$}
\put(4.9,6.6){${\cal H}^-$}
\put(-3.5,4.3){(a)}
\put(3.5,4.3){(b)}
\end{picture}
\label{f:bbh}
\end{center}
\end{figure}

In \sec{s:confdual}, we review the qualitative properties of the dual gravity solutions as a function of  $HR$. For small $HR$ the dual gravity solution is the well known ``bubble of nothing'' in AdS (see \fig{f:bbh}(a)) \cite{Aharony:2002cx}. These spacetimes have been explored in the past in several works \cite{Balasubramanian:2002am, Balasubramanian:2005bg,He:2007ji} in the context of the AdS/CFT correspondence. In the present discussion, the appearance of a ``bubble'' is very natural: it corresponds to the fact that the gravity solutions dual to confining gauge theories smoothly cap-off in the the infra-red (IR). For a dS boundary geometry, the IR ``end'' should also look like a de Sitter space; this IR end is the expanding bubble wall.\footnote{In various earlier works, it has been suggested that there may be non-geometrical degrees of freedom associated with the ``nothing'' inside the bubble; from our present perspective, this seems unnecessary; the presence of a bubble is directly related to the fact that the corresponding field theory has a mass gap.} For large $HR$ we argue following \cite{Hutasoit:2009xy} that  the relevant bulk solution is  the ``topological black hole'' described by Ba\~nados \cite{Banados:1997df}  whose causal diagram is shown in \fig{f:bbh}(b). This solution has a horizon in the bulk, in accord with expectations for the dual to be a deconfined plasma phase.

While the heuristic description given above is indicative of a confinement/deconfinement transition as a function of $HR$, one would also like to verify this more directly. To this end we review in \sec{s:ordpar} that certain conventional order parameters for confined and deconfined phases of the gauge theory in Minkowski spacetime can be ported over to the dS analysis and serve to distinguish the low-$HR$ and high-$HR$ phases. We further argue  that there should exist gravity solutions corresponding to the field theory in the static patch of dS spacetime with arbitrary $T$ and $H$, so there is a natural phase diagram parameterized by $T$ and $H$, for fixed $R$. On this diagram, the low $T$ Minkowski solution and the low $H$ de Sitter solution may be smoothly deformed into one another, so both of these belong to a single generalized confining phase. Finding explicit dual gravity solutions which smoothly interpolate between the high $H$ de-Sitter phase and the large $T$ Minkowski phase would demonstrate definitively that both of these belong to the same generalized deconfined phase, but the best we are able to do is is give an implicit construction of the interpolating solution at leading order in $H$ (about Minkowski spacetime).

Finally, in \sec{s:discuss} we offer some concluding remarks and point out some interesting questions for future research.

\paragraph{Relation to previous work:} Before proceeding to describe our set-up, it is useful to recall some of the efforts of trying to understand the dynamics of field theories in de Sitter spacetime using the AdS/CFT correspondence.  For instance the holographic duality between the \AdS{5}  bubble of nothing spacetime \cite{Aharony:2002cx} and field theories on \dS{3} $\times \,\Sp^1$ was first discussed in  in \cite{Balasubramanian:2002am}.\footnote{A large class of bubble geometries were also constructed in \cite{Cvetic:2003zy}.} These authors pointed out that the boundary of the bubble of nothing is \dS{3} $\times \,\Sp^1$, thereby allowing one to examine the dynamics of strongly coupled theories on \dS{3} and computed the boundary stress tensor. Subsequently, \cite{Cai:2002mr} noted that the topological AdS black hole of \cite{Banados:1997df} also has the same boundary. These disparate pieces were put together in \cite{Ross:2004cb} who used the two distinct bulk solutions to investigate the behaviour of the $\alpha$-vacua in de Sitter spacetime for strongly coupled theories. In particular, it was argued that the $\alpha$-vacua are unphysical since they would require additional singularities at anti-podally related points on the black hole horizon for Green's functions.

The \AdS{5} bubble of nothing and the topological black hole spacetimes were further investigated in \cite{Balasubramanian:2005bg,He:2007ji} with an aim towards understanding the vacuum stability. It was argued that the bubbles of nothing mediate  a semi-classical decay of the topological black hole spacetime, for small enough temperatures. More recently, \cite{Hutasoit:2009xy} interpreted the bubble of nothing and the Ba\~nados black hole as two different phases of the de Sitter field theory, noting that the bubble of nothing has lower free energy at small $HR$ and that the Ba\~nados black hole dominates at large $HR$; indeed, the bubble of nothing solutions exist only for $HR$ below a certain critical upper bound.  The fact that the Scherk-Schwarz circle pinches off in the bubble of nothing was used to argue that the associated Wilson loops have a finite expectation value in this phase and that the ${\mathbb Z}_N$ symmetry of the dual gauge theory is broken.  On the other hand, the Ba\~nados black hole phase in ${\mathbb Z}_N$ invariant. By exploring the real-time behavior of Green's functions, \cite{Hutasoit:2009xy} provided evidence that the bubble of nothing phase is characterized by gapped hadronic excitations, while the Ba\~nados black hole phase should be interpreted as a plasma.

We also mention various attempts to holographically understand the behaviour of inflationary dynamics in terms of quantum field theories, see
\cite{Hertog:2004rz,Hertog:2005hu}, \cite{Freivogel:2005qh}, \cite{Craps:2005wd}, \cite{Chu:2006pa,Das:2006dz}, \cite{Freivogel:2006xu}, \cite{Buchel:2002wf,Hirayama:2006jn} for a sampling of ideas related to this effort. We should note that while it would be fascinating to have a holographic dual of an eternally inflating geometry, and moreover is in fact necessary in a quantum gravitational context,  our goals are more modest. We simply wish to ask about the physical behaviour of a field theory with non-trivial phase structure on a fixed non-dynamical background geometry which we take to be of the inflationary cosmology. Closer in spirit to the current undertaking is the recent work of \cite{Chesler:2008hg,Bhattacharyya:2009uu} where the authors considered the holographic dual of a field theory in a time-varying background (which in these works was not a cosmology, rather the time dependence was engineered to inject energy into the system).

\section{Field theory in de Sitter: a brief review}
\label{s:bkgnd}

To set the stage for our discussion  we review a few basic properties of de Sitter spacetime describing the various coordinate charts we employ. Furthermore, we will discuss the salient aspects of QFT on dS with specific focus on thermal physics in the static patch. A more complete recent review and further references may be found in \cite{Spradlin:2001pw}.
\subsection{The de Sitter geometry}
\label{s:dsbasics}
The dS spacetime in $d$ dimensions is the maximally symmetric solution to Einstein's equations with positive cosmological constant. It is most simply described in terms of its embedding as a hyperboloid
\begin{equation}
X_i^2 - X_{0}^2 = R_{\text dS}^2  = \frac{1}{H^2} \ ,
\label{dshyp}
\end{equation}
in ${\bf R}^{d,1}$. The metric on the spacetime can be obtained by pulling back the standard flat Minkowski metric on ${\bf R}^{d,1}$ onto the hyperboloid.

One can define  global coordinates that cover the entire spacetime. These coordinates make manifest the  $SO(d) \times {\bf R} \subset SO(d,1)$ isometry  and the metric may be written as
\begin{equation}
ds^2 = -d\tau^2 + {1 \over H^2} \cosh^2 \left(H\, \tau\right) \, d\Omega_{d-1}^2\ .
\end{equation}

For our purposes, it will be useful to consider a couple of other coordinate systems. First, we note that dS spacetime can be written in FRW form as
\begin{equation}
ds^2 = -dT^2 + e^{2 H\,  T} \, d{\bf x}^2 \ , \qquad {\bf x} \in {\bf R}^{d-1}
\end{equation}
In this form, the spatial sections are flat, and the metric provides a good approximation to the presumed spacetime geometry during an inflationary phase of early-universe cosmology. These coordinates however cover only part of the spacetime; in particular they cover the causal past of an observer sitting at the north pole of $\Sp^{d-1}$.

Finally, we can choose static coordinates which cover the causal diamond associated with a geodesic observer. For convenience, we pick the observer to be the one sitting at the north pole of the $\Sp^{d-1}$ and write the metric in static, spherically symmetric form:
\begin{equation}
ds^2 = -(1-H^2 \,r^2)\, dt^2 + \frac{dr^2}{1 -H^2 \,r^2} + r^2 \, d \Omega_{d-2}^2 \; .
\end{equation}
Here $r=0$ corresponds to the north pole of the $\Sp^{d-1}$ and the cosmological horizon is located at $r = H^{-1}$, see for instance \fig{f:staticp}.  With this choice of coordinates, the spatially-integrated energy density $T^{00}$ defines a conserved energy, which will form the basis for our discussion of thermodynamics in de Sitter spacetime.

\subsection{Quantum field theory on de Sitter spacetime}
\label{s:qftds}

The de Sitter spacetime has a high degree of symmetry: \dS{d} is maximally symmetric with a $SO(d,1)$ isometry group. When one considers a local quantum field theory on this background, it is therefore natural to ask for a vacuum state which respects this symmetry. Such a state is said to be de Sitter-invariant. For a free field theory, it turns out that there is a one-parameter family of de-Sitter invariant vacuum states, known as the $\alpha$-vacua \cite{Mottola:1984ar,Allen:1985ux}. Among these, there is a canonical choice (the Euclidian or Bunch-Davies vacuum \cite{Bunch:1978yq}) that reduces to the standard Minkowski space vacuum state in the limit $H \to 0$. The correlation functions in this vacuum are related by analytic continuation to the correlation functions for the Euclidean field theory on the round sphere $\Sp^d$. Henceforth, when referring to the de-Sitter invariant vacuum state, we will mean the Euclidean vacuum.\footnote{There is some debate about whether the more general $\alpha$-vacua can be consistently defined in the presence of interactions. If $\alpha$-vacua exist for strongly coupled CFTs with a gravity dual, there should be a one-parameter family of gravity solutions dual to these states. It would be very interesting to determine   conclusively whether these exist, but it seems difficult to find a consistent bulk theory which generates the antipodal singularities of correlation functions that are characteristic of $\alpha$-vacua \cite{Ross:2004cb}. An attempt to address such issues was made in \cite{Hutasoit:2009sc}, though the relation to $\alpha$-vacua is unclear.}.

The Euclidean vacuum in de Sitter spacetime may be characterized by a temperature $T_{dS}$ related to the Hubble parameter by \cite{Gibbons:1977mu}
\begin{equation}
T_{\text dS} = \frac{1}{2\pi} \, H \equiv \frac{1}{2\pi\,R_{dS}} \; .
\label{}
\end{equation}
Physically, an observer on a timelike geodesic equipped with an Unruh detector will observe a thermal bath of particles at this temperature.

Note that the de Sitter invariant vacuum state is well defined on the entire manifold. One can choose to view this state in the static patch, where a local observer would associate the thermal nature of the vacuum to the presence of his/her cosmological horizon. This is similar to the Hartle-Hawking state of black holes and describes the sensible equilibrium state in the geometry. In particular, it is worth noting that the state is regular on the cosmological horizon (this follows from the Euclidean continuation).

\subsection{Thermodynamics in the static patch of de Sitter Space}
\label{s:staticp}

As we have reviewed, quantum field theory in the de Sitter-invariant vacuum state has a characteristic temperature inversely proportional to the de Sitter radius. However, in the static patch of dS, it is also possible to study field theory at other temperatures, as we now explain.

 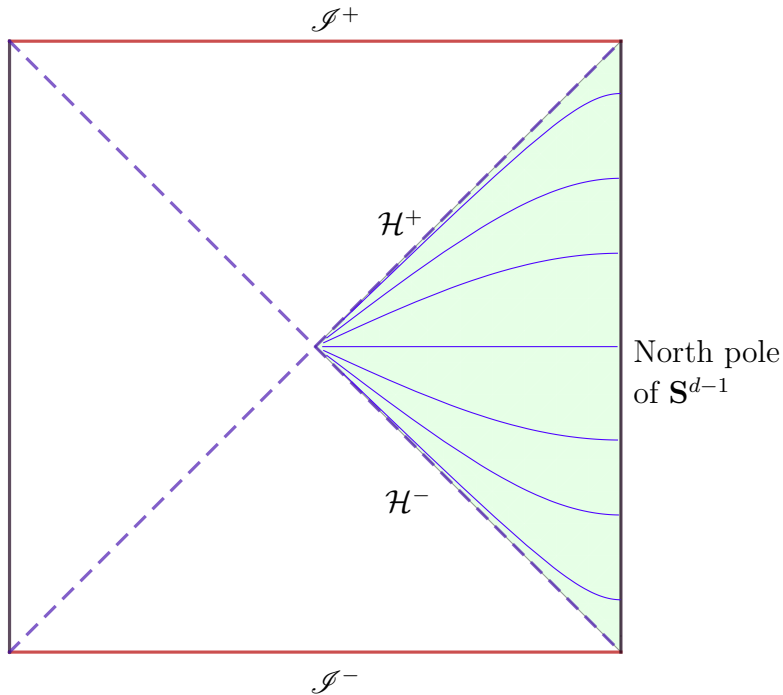
\begin{figure}[t]
\begin{center}
\includegraphics[scale=1]{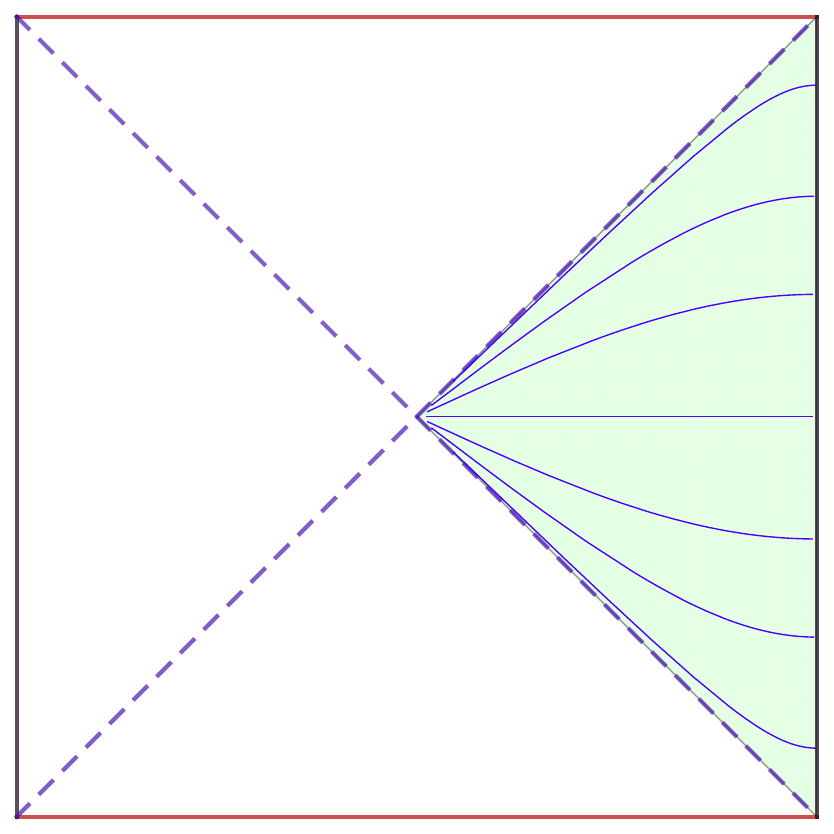}
 \caption{Penrose digram for de Sitter spacetime in $d$ dimensions. The  static patch associated with the observer on the north pole of the $\Sp^{d-1}$ is shown as the shaded region.  Lines of constant $t$ are  indicated  as the solid (blue) curves. The dashed lines are the cosmological horizons, with $ {\cal H}^\pm$ being the future/past cosmological horizons of the static observer.}
\begin{picture}(0,0)
\setlength{\unitlength}{1cm}
\put(0,11.2){${\mathscr I}^+$}
\put(0,2.4){${\mathscr I}^-$}
\put(0.9, 8.5){${\cal H}^+$}
\put(1,4.8){${\cal H}^-$}
\put(4.3, 6.8){North pole}
\put(4.3,6.25){of $\Sp^{d-1}$}
\end{picture}
\label{f:staticp}
\end{center}
\end{figure}

In the static patch description of dS we have a timelike Killing vector $\partial_t$, and thus a conserved energy (which can be defined by integrating $T^{00}$ over a surface of constant $t$, see \fig{f:staticp}).  This is all we need to discuss the usual thermodynamic ensembles for quantum field theory as long as we restrict attention to the static patch. In particular, we can ask about the equilibrium state in the canonical ensemble for any given $H$ and any temperature $T$ which may not be equal to the de Sitter temperature; this should be the state of minimum free energy. For any field theory, we can draw a phase diagram as a function of these two variables.

To understand the issues involved here a bit better, it is useful to pass to the Euclidean geometry. Analytically continuing $t \to -i \,t_E$ and redefining $r = H^{-1} \,\sin \theta$ one has the standard round metric on $\Sp^d$, with $t_E$ now being identified as one of the angular isometries:
\begin{equation}
\label{geom}
ds^2  = \frac{1}{H^2} \, \left(H^2\, \cos^2\theta\,  d t_E^2 + d \theta^2 + \sin^2\theta\,  d \Omega_{d-2}^2\right) \; .
\end{equation}
This  geometry is non-singular only if we take the period of $t_E$ to be $\frac{2 \pi}{H}$ and thus the state of the Euclidean QFT to be at $T=T_{\text dS}$. We can nevertheless pick a different periodicity for $t_E$, say $t_E \sim t_E + \frac{1}{T}$  which corresponds to a temperature $T \neq T_{\text dS}$. We denote such geometries as $\Sp^d_\chi$ characterized by the dimensionless ratio $\chi \equiv  \frac{H}{T}$. This Euclidean manifold has singularities at $\theta = \pm \frac{\pi}{2}$ which map in the Lorentzian spacetime to the de Sitter horizon. As long as we do not try to extend the geometry through the de Sitter horizon, there is nothing pathological about the associated Lorentzian spacetime. Physically, we can imagine that there is a heat bath sitting at, or just in front of the de Sitter horizon in order to hold the temperature at a temperature different from $T_{dS}$. In this picture, the geometry does not smoothly continue through the horizon because we run into the heat bath which is visible as the conical singularity in $\Sp^d_\chi$.

If we were considering perturbative quantum fields, we would perform the Euclidean path integral for the field theory on $\Sp^1_\chi$ to compute say local correlation functions.
On the other hand, for theories with a gravity dual description, the equilibrium state of the field theory for general $T$ and $H$ will be determined by the minimum action bulk solution of the appropriate gravity theory with boundary geometry $\Sp^d_\chi$. It is important to note that despite the conical singularity in the boundary geometry for $\chi \ne 2 \pi$, the bulk geometry can be completely non-singular.\footnote{To illustrate this, consider an (American) football-shaped region of flat Euclidean space. The interior of the football is clearly non-singular, but the boundary has conical singularities at the ends of the football.}

\section{Holographic duals of CFTs on de Sitter spacetime}
\label{s:cftds}

As described in the Introduction, we wish to investigate strongly interacting quantum fields in de Sitter spacetime. While generic theories are hard to understand in the strong coupling limit, there exists a class of theories which can be understood via the AdS/CFT correspondence.  While we will specifically have such holographic models in mind, we hope that the general lessons extracted here will continue to hold in the case of physically interesting theories.

The essence of the AdS/CFT correspondence is that the strongly coupled field theory dynamics is recorded in terms of string theory (or classical gravity if the field theory admits an appropriate planar limit) with appropriate asymptotically Anti-de Sitter boundary conditions. If the bulk AdS spacetime geometry is some negatively curved $(d+1)$-dimensional Lorentzian manifold, ${\cal M}_{d+1}$, with conformal boundary $\partial {\cal M}_{d}$, then the  field theory lives on a spacetime ${\cal B}_d$ of dimension $d$ in the same conformal class as $\partial {\cal M}_{d}$. Choosing an appropriate conformal frame, one may identify ${\cal B}_d$ and $\partial {\cal M}_{d}$ and speak of the field theory as living on the AdS boundary.  From the standpoint of the bulk theory, the choice of metric on ${\cal B}_d$ fixes a boundary condition that the bulk solution must satisfy.\footnote{In standard AdS/CFT parlance, this amounts to fixing the non-normalizable mode of the bulk graviton to obtain the desired metric on ${\cal B}_d$.} The correspondence is simplest to state for conformal field theories in dimensions where the trace anomaly vanishes, but with appropriate care the correspondence also holds in the presence of a trace anomaly and it can accommodate non-conformal deformations.

Given this set-up  we can probe the detailed phase structure of strongly coupled CFTs  on de Sitter spacetime simply by taking the boundary manifold $ {\cal B} =$ \dS{d}.   The space of field theory states on  \dS{d} then is simply the same as the classical solution space of asymptotically locally AdS states of string theory subject to the  boundary condition above.  In the planar limit of the field theory, the latter is essentially the solution space of classical (super-)gravity with the given boundary condition.  The different phases of the CFT on \dS{d} should thus correspond to different bulk spacetime geometries ${\cal M}_{d+1}$ in this solution space.

Let us therefore consider a CFT in $d$-dimensions which we know has a holographic dual in terms of gravity in \AdS{d+1} spacetime. Prototypical examples of such CFTs are the ${\cal N} =4$ Super Yang-Mills theory or other ${\cal N} =1$ superconformal field theories in four dimensions obtained by placing D3-branes at various singularities. Likewise examples in other dimensions can be constructed by placing other brane configurations available in string/M-theory at appropriate singularities. For all such examples the conformally invariant vacuum state of the CFT on \ESU{d} $\equiv {\bf R} \times \Sp^{d-1}$ is just dual to the global \AdS{d+1} spacetime.

To investigate the CFT dynamics on \dS{d} it suffices to consider the universal sector of the AdS/CFT correspondence, wherein the bulk spacetime ${\cal M}_{d+1}$ is a solution to Einstein's equations with a negative cosmological constant. The bulk action is therefore taken to be\footnote{We will use $g_{AB}$ to denote the bulk metric on ${\cal M}_{d+1}$ with uppercase Latin indices indicating the bulk  spacetime dimensions. The metric on ${\cal B}_d$ will be denoted as  $\gamma_{\mu \nu}$ with lowercase Greek indices labeling boundary directions.}
\begin{equation}
{\cal S}_{\text{bulk}}  = \frac{1}{16 \pi \, G^{(d+1)}_N}\,\int d^{d+1}x \, \sqrt{-g}\, \left( R - 2 \, \Lambda\right)  \ ,
\label{bulkact}
\end{equation}
where we can express the cosmological constant as $\Lambda = -\frac{d\,(d-1)}{2}\,\ell_{d+1}^{-2}$ in terms of the AdS$_{d+1}$ radius $\ell_{d+1}$.

The task at hand is therefore finding all possible geometries which have as their boundary \dS{d}. As explained earlier we are not only interested in the case where the Euclidean boundary manifold is smooth, but also in the situation where we allow for a temperature differing from $T_{\text dS}$ by restricting attention to the static patch. Our analysis will proceed in two stages: first we discuss the rather simple case of 1+1 dimensional CFTs on \dS{2} and then move onto higher dimensional examples.  Since we allow $T\neq T_{\text dS}$, there is in principle the possibility of phase transitions determined by the value of $T/T_{\text dS}$.  However, we find no such phase transitions below, even at $T/T_{\text dS}=0$.

\subsection{Conformal field theory on \dS{2}}
\label{s:2dds}

We begin our discussion with the behaviour of a  1+1 dimensional CFT  on \dS{2}. The CFT in question is assumed to have a weakly curved gravity dual, which at the very least necessitates that the central charge $c \gg 1$.\footnote{In fact in all known examples not only is the central charge large, but there is a large gap in the spectrum of conformal dimensions. It has been recently been argued in \cite{Heemskerk:2009pn} that these two constraints might even be a sufficient condition for the existence of a holographic dual.} Prototype examples of such theories are the world-volume theories on a bound state of D1 and D5 branes, where the central charge $c \sim \sqrt{Q_1\,Q_5}$ is large when the number of branes $Q_{1,5} \gg 1$.

Let us first consider the situation where we have the global \dS{2} spacetime on which the CFT resides.  As argued earlier, the theory has an unique de Sitter invariant vacuum. This is in fact easy to see holographically; the unique Euclidean geometry whose boundary is the round $\Sp^2$ is the Euclidean \AdS{3} spacetime. One simply analytically continues the Euclidean time coordinate both on the boundary and in the bulk. In essence, the holographic dual for a 1+1 CFT on \dS{2} is the \AdS{3} spacetime written in a \dS{2} foliation:
\begin{equation}
ds^2 =d\rho^2 + (H\, \ell_3)^2 \, \sinh^2\left(\frac{\rho}{\ell_3}\right)  \left(-d\tau^2  + {1 \over H^2} \cosh^2( H\, \tau)\, d\phi^2\right)
\label{adsinds}
\end{equation}
The Euclidean geometry with $\tau \to i\, t_E + \frac{2\pi}{ H}$ is unique, as it should be since one has a unique conformally invariant vacuum state for the field theory.

One can equivalently consider the static patch patch of \dS{2}: this can be achieved by simply making a coordinate transform in \req{adsinds}. The bulk is then an \AdS{3} spacetime in a warped product form, where the leaves of the foliation are static \dS{2} geometries. This is appropriate for a static observer who experiences a temperature $T = T_{\text dS}$.

We would now like to find gravity solutions dual to the CFT on the static patch of de Sitter spacetime with Hubble parameter $H$ and and arbitrary temperature $T$. Since the bulk theory is three-dimensional, the dual gravity solution must be locally \AdS{3} everywhere. In particular, this means that the desired geometry should be obtained simply by a coordinate transformation of global \AdS{3}. One can equivalently employ the trick described in \cite{Hubeny:2009ru} to construct the geometry in a perturbation expansion in Fefferman-Graham coordinates. We relegate the details of this construction to \App{s:ds2sol} and simply quote the final solution here. The metric takes the form:\footnote{In \cite{Hubeny:2009ru} this trick was used to construct the so called black funnel geometry in \AdS{3} where the boundary metric was a two dimensional black hole spacetime.}
\begin{equation}
ds^2 = {\ell_3^2 \over z^2} (- f(r,z) dt^2 + j(r,z)dr^2 + dz^2 ) \;.
\label{ds2HT}
\end{equation}
where
\begin{eqnarray}
j(r,z) &=& {1 \over 4} \left({1 \over \sqrt{1 - H^2\, r^2}} + z^2 \left\{ -  {H^2  \over \sqrt{1- H^2\, r^2}} + {(2 \pi\, T)^2 - H^2 \over (1- H^2 \,r^2)^{3 \over 2}} \right\}\right)^2 \cr
f(r,z) &=& {1 \over 4} \left(\sqrt{1 - H^2\, r^2} + z^2 \left\{ -  H^2 \sqrt{1 - H^2\, r^2} -  {(2 \pi\, T)^2 - H^2  \over \sqrt{1- H^2\, r^2}} \right\} \right)^2
\label{dsfinal}
\end{eqnarray}
Note that we have switched to using Fefferman-Graham coordinates. For the case $H = 2 \pi T$, the metric \req{ds2HT} is related to (\ref{adsinds}) by simply changing to static coordinates in the de Sitter factor and defining $z = H^{-1}e^{-\frac{\rho}{\ell_3}}$, making it clear that the boundary of  \AdS{3} is obtained as $z \to 0$.  For other values of $T$, the metric \req{ds2HT} again describes a patch of pure \AdS{3}, but now with a different choice of conformal frame at infinity.

\begin{figure}[t]
\begin{center}
\includegraphics[scale=1]{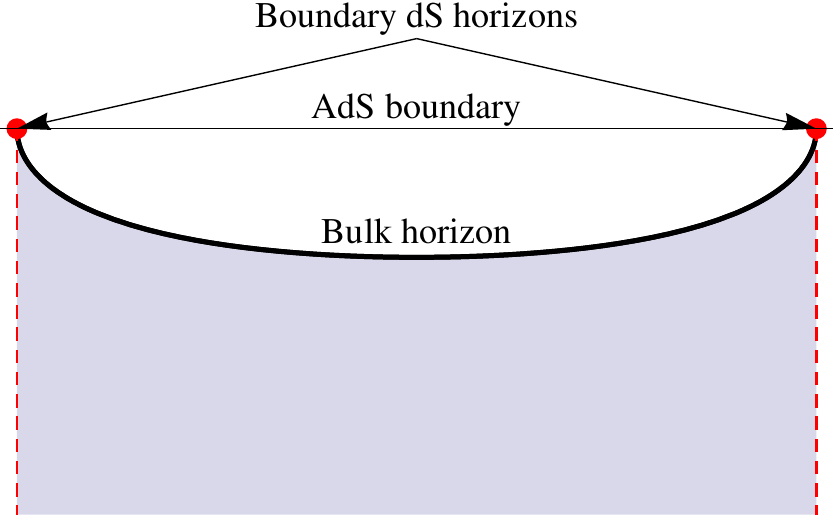}
 \caption{The plot of the bulk horizon for the bulk three dimensional geometry \req{ds2HT} whose boundary is \dS{2}.  We plot the location of the horizon in the $\{r,z\}$ coordinates used to express the solution as a solid black line. The static patch of de Sitter is restricted to $|r| \le H^{-1}$ and this is the relevant region of the spacetime even in the bulk.  The shaded region lies inside the horizon in the bulk geometry.}
\label{f:ds2btz}
\end{center}
\end{figure}

The solution \req{ds2HT}, \req{dsfinal} has a horizon at
\begin{equation}
z = \sqrt{\frac{1-H^2\,r^2}{(2 \pi \,T)^2 - H^4\,r^2}}
\label{bulkhor2}
\end{equation}
in addition to a horizon at
\begin{equation}
r = \frac{1}{H} \ , \qquad\forall z
\label{bdyhor2}
\end{equation}
The horizon \req{bulkhor2} asymptotes to the horizon \req{bdyhor2} as indeed it must. One can however argue that the horizon \req{bulkhor2} is the relevant horizon in the bulk.  On this one can check that the surface gravity is  constant, and that the bulk horizon is smooth. The temperature is thus fixed and we have chosen normalizations so that it is given by $T$, independent of $H$, as we hoped to achieve with our construction.\footnote{In the context of  \AdS{4} C-metrics one can similarly construct the so called black droplet geometries \cite{Hubeny:2009kz} which bear qualitative resemblance to the solution considered here. In particular, there are solutions where the boundary resembles a black hole in a closed universe which have bulk horizons that resemble \req{bulkhor2}. } A plot of the bulk horizon is shown in \fig{f:ds2btz}.

Independent of the explicit construction sketched in \App{s:ds2sol} we can  argue as follows that such a solution must exist.  The static patch of \dS{2} (for any value of $H$) is conformally related to Minkowski space $ {\bf R}^{1,1}$ as both are (conformally) diamonds.  In fact, since Minkowski space is scale-invariant, it is clear that we can take them to be conformally related by a conformal factor that is a) independent of time $t$ and b) equal to 1 at $r=0$ for all $t$.  Such a conformal transformation maps the timelike Killing field $\partial_t$ of Minkowski space  to the usual static patch Killing field $ \partial_t$ of \dS{2} (with the standard normalization).

Suppose one starts with \AdS{3} in the BTZ coordinates, where the boundary is  (two copies of) Minkowski space and the solution is periodic in Euclidean time with period $T^{-1}$ (which is related to the BTZ parameter).  The above mentioned change of conformal frame maps ${\bf R}^{1,1}$ to the static patch of \dS{2}, but since it  maps the timelike Killing fields properly $\left(\partial_t \right)_{\text Minkowski} \to \left(\partial_t\right)_{\text dS}$, it follows that the Euclidean period still remains  $T^{-1}$.  One thus obtains a  solution with boundary Hubble constant $H$ and temperature $T$. Though the construction is slightly different in form, this solution coincides with the one derived differently in \App{s:ds2sol}.

Given the explicit solution \req{ds2HT}, \req{dsfinal} it is easy to evaluate the expectation value of the stress tensor. This is done by computing the quasi-local Brown-York stress tensor using the standard AdS/CFT dictionary \cite{Balasubramanian:1999re,Henningson:1998gx}. For our purposes it is convenient to derive the boundary stress tensor for the configuration using the Fefferman-Graham expansion. The bulk metric $g_{AB}$ is already in the Fefferman-Graham gauge; all we need to do is to expand the metric at a constant $z$ slice, call it $\Gamma_{\mu\nu}$, in  a  Taylor series expansion in $z$ with coefficients $\Gamma_{(k)}$. Given this expansion say in the form
\begin{equation}
\Gamma (x,z) = \Gamma_{(0)}(x) + z^2\, \Gamma_{(2)} + \cdots
\label{Gamexp}
\end{equation}
the holographic stress tensor is \cite{deHaro:2000xn}:
\begin{equation}
 T_{\mu \nu} =  c\, \left(\Gamma_{(2) \mu \nu} - \Gamma_{(0)\mu \nu}\, \Tr{\Gamma_{(2)}}\right) \ .
\label{holsten}
\end{equation}
where of course $\Gamma_{(0)}$ is the metric $\gamma$ on the boundary and $c = \frac{2  \,\ell_3}{16\pi\, G^{(3)}_N}$ is proportional the boundary central charge.

For \eqref{dsfinal} one finds (separating the trace free and trace parts of the stress tensor):
\begin{equation}
T^{\;\mu}_\nu =c\, H^2 \left[\frac{{\cal F}_2(T,H) }{1-H^2\,r^2} \;  \text{diag}\big\{-1,1 \big\} +  \text{diag} \big\{1,1 \big\} \right] ,
\label{}
\end{equation}
where for future convenience we have defined
\begin{equation}
{\cal F}_d(T,H) = \left(\frac{4\pi\, T}{d\, H}\right)^{d-2} \, \left(\left(\frac{4\pi\, T}{d\, H}\right)^{2} -1 \right) .
\label{Fd}
\end{equation}
This stress tensor is not regular at the boundary \dS{2} horizon at $r = H^{-1}$, unless $2 \pi \,T = H$, i.e. unless we are in the de Sitter vacuum state. The lack of regularity for other temperatures is expected, since for these temperatures, we must have a heat bath at the de Sitter horizon. Also note that the stress tensor is not traceless; this is simply a manifestation of the conformal anomaly in 1+1 dimensions. In addition it is also useful to see that the result \req{hosten} nicely interpolates between the stress tensor on de Sitter at $T_\text{dS}$ and the result for a theory in Minkowski space at temperature $T$. Finally, we note that our solution is unique for all values of $T \ge 0$, indicating that there are no phase transitions for any value of $T/T_{\text dS}$.

\subsection{Higher-dimensional conformal field theories on de Sitter spacetime.}
\label{s:hdcft}

Having discussed the simple example of 1+1 dimensional CFTs, we now turn to higher dimensional CFTs in \dS{d} spacetimes. Once again it is easy to write down the holographic dual for the CFT on global de Sitter spacetime. The Euclidean geometry is again uniquely determined to be \AdS{d+1} in order to maintain full $SO(d,1)$ invariance. The bulk spacetime metric takes the form:
\begin{equation}
ds^2 =d\rho^2 + (H\, \ell_{d+1})^2\, \sinh^2\left(\frac{\rho}{\ell_{d+1}}\right)  \left(-d\tau^2  + {1 \over H^2}\cosh^2( H\, \tau)\, d\Omega_{d-1}^2\right)
\label{adsindsd}
\end{equation}
One can again transform this into the static coordinates and recover the holographic dual for CFTs at $ T_{\text dS}$.

In order to consider CFTs at temperatures differing from $T_{\text dS}$ one needs a bit more work.
The situation in a 2+1 dimensional bulk was somewhat special in that all negatively curved Einstein three manifolds are diffeomorphic to \AdS{3}. Therefore the problem of finding three dimensional bulk geometries with specified 1+1 dimensional boundaries reduces to finding an appropriate diffeomorphism, which is a simpler (albeit not necessarily trivial) problem. For higher dimensional examples one does not have this luxury. A-priori it appears as though one has to solve a co-homogeniety two system of partial differential equations to determine the appropriate bulk geometry, since the boundary \dS{d} metric has only manifest $ {\bf R}\times SO(d)$ isometry.

Happily, this complication can be circumvented and the desired bulk spacetimes can be found explicitly using the Weyl rescaling idea sketched in \sec{s:2dds}. The basic idea is the following: we note that the static patch of \dS{d} is conformal to the Lorentzian hyperbolic cylinder ${\bf R} \times {\bf H}_{d-1}$ where ${\bf H}_{d-1}$ is the Euclidean hyperboloid (i.e., Euclidean AdS). Moreover, the static Killing field $\partial_t$ of de Sitter maps to the static Killing field of the hyperbolic cylinder as can be see  by writing:
\begin{equation}
ds^2 = (1-H^2\,r^2) \, \left(-dt^2 + \frac{dr^2}{(1-H^2\, r^2)^2} + \frac{r^2}{1-H^2\,r^2}\, d\Omega_{d-2}^2\right)
\label{confhc}
\end{equation}
and performing the coordinate change $H\,r = \tanh \xi$. Therefore, if one has the holographic duals for the hyperbolic cylinder with Euclidean time period given by $T^{-1}$ with $ T \neq T_{\text dS}$, then by the above described Weyl rescaling one can immediately construct the duals for the static patch of \dS{d} with $T \neq T_{\text dS}$ (or equivalently for the boundary Euclidean manifold $\Sp^1_\chi$).

Fortunately, such solutions are indeed known; these are the hyperbolic (also referred to sometimes as topological) black holes described in \cite{Emparan:1998pf,Birmingham:1998nr,Emparan:1999gf}. In addition, it was argued in \cite{Emparan:1998pf,Birmingham:1998nr} that these are the unique solutions with the desired symmetries and thus that there are no phase transitions for the dual gauge theory on ${\bf R} \times {\bf H}_{d-1}$.  It follows that there are again no phase transitions on de Sitter as a function of $T/T_{\text dS}$.

The metric for the hyperbolic black hole geometries takes the form:
\begin{equation}
ds^2 = -f(\rho) \, dt^2 + \frac{d\rho^2}{f(\rho)} + \rho^2 \, d\Sigma_{d-1}^2  \ , \qquad f(\rho ) = \frac{\rho^2}{\ell_{d+1}^2} - 1 - \frac{\rho_+^{d-2}}{\rho^{d-2}}\left(  \frac{\rho_+^2}{\ell_{d+1}^2}-1\right)
\label{hypbh}
\end{equation}
where $d\Sigma_{d-1}^2 = d\xi^2 + \sinh^2 \xi\, d\Omega_{d-2}^2$ is the metric on the Euclidean hyperboloid. This bulk $d+1$ dimensional spacetime has a horizon at the zero locus of $f(\rho)$ and the corresponding Euclidean solution will be regular provided
\begin{equation}
t_E \sim t_E + %
\beta \ , \qquad %
\beta \equiv T_\text{hyp}^{-1} = \frac{4\pi\, \ell_{d+1}^2 \, \rho_+}{d\, \rho_+^2 - (d-2)\, \ell_{d+1}^2}.
\label{hyptemp}
\end{equation}

Given this hyperbolic black hole \req{hypbh}, all we need to do is perform a diffeomorphism in the bulk that acts as a boundary conformal transformation, achieving the mapping indicated in \req{confhc}. This is easily done for instance by defining
\begin{equation}
r = \frac{1}{H} \, \tanh\xi
\label{}
\end{equation}
which results in the metric
\begin{equation}
ds^2 = \frac{H^2\, \rho^2}{1-H^2\,r^2} \, \left(-\frac{f(\rho)}{H^2\,\rho^2}\, (1 -H^2\,r^2)\, dt^2 + \frac{dr^2}{1-H^2\,r^2} + r^2\, d\Omega_{d-2}^2 \right) + \frac{d\rho^2}{f(\rho)}.
\label{dsdduals}
\end{equation}
We claim that \req{dsdduals} are the bulk geometries appropriate to describe the dynamics of CFTs on the static patch of \dS{d} at temperature $T$ given by \req{hyptemp}, which in general differs from the de Sitter temperature $T_\text{dS}$. Of course, for the particular case $T=T_\text{dS}$, the metric \req{dsdduals} differs from \req{ds2HT} only by a simple coordinate change.

Having identified the relevant holographic dual geometries we are in a position to answer physical questions about the dynamics of CFTs at $T \neq T_\text{dS}$ on the static patch of \dS{d}. The simplest observable one can extract from the holographic dual is the quantum expectation value of the stress tensor. As described in \sec{s:2dds} one can compute this by using the holographic prescription of \cite{Balasubramanian:1999re,Henningson:1998gx}. Fortunately, our task is simplified for the case of odd dimensional de Sitter spacetimes ($d$ odd) since one can obtain the result by following the conformal transformation.

First of all note that the boundary stress tensor for the hyperbolic black hole \req{hypbh} has been previously determined in \cite{Emparan:1999gf}:
\begin{equation}
T_\nu^{\; \mu} = \frac{1}{16\, \pi\, G_N^{(d+1)}\, \ell_{d+1}} \left(\varepsilon + \frac{\rho_+^{d-2}}{\ell_{d+1}^{d-2}} \left(\frac{\rho_+^{2}}{\ell_{d+1}^{2}} -1\right) \right) \text{diag}\bigg\{1-d,1,1,\cdots,1\bigg\}
\label{}
\end{equation}
 where
\begin{equation}
\varepsilon =  \frac{(d-1)!!^2}{d!}  \qquad \text{for\; even\;} d \ ,
\label{}
\end{equation}
and vanishes for odd $d$. We then note that under a Weyl rescaling of the metric $g_{\mu\nu} = e^{2\phi} \, {\tilde g}_{\mu\nu}$ the stress tensor transforms as $T^{\;\mu}_\nu  = e^{-d\,\phi}\, {\tilde T}^{\;\mu}_{\nu}$. This suffices to explicitly determine the boundary stress tensor for CFTs on \dS{d} for odd $d$ -- in particular,  we find the following stress tensor on the static patch of de Sitter with temperature $T$ in terms of the central charge $c = \frac{\ell_{d+1}^{d-1}}{16\pi\,G^{(d+1)}_N}$) of the CFT:\footnote{We have written the final expression for the stress tensor in terms of the field theory temperature $T$ and the Hubble scale $H$. The field theory temperature is related to the temperature of the hyperbolic black hole via $T = H\, \ell_{d+1}\, T_\text{hyp}$ as can be inferred by tracking the change in conformal frame.}
\begin{equation}
T_\nu^{\; \mu} =  c\, H^d\, \frac{{\cal F}_d(T,H)}{(1-H^2\,r^2)^{\frac{d}{2}}}\;  \text{diag}\bigg\{1-d,1,1,\cdots,1\bigg\}
\label{Todddds}
\end{equation}
where ${\cal F}_d$ is defined in \ref{Fd}.

The expression for the one-point function of the stress tensor \req{Todddds} is accurate for odd $d$ because  there is no conformal anomaly in odd dimensions. For even dimensional CFTs, the above expression does not capture the conformal anomaly. One then indeed has to use the covariant formalism of boundary stress tensor or transform to Fefferman-Graham coordinates to extract the expectation value of the stress tensor. For example in $d=4$ one would find a coordinate transformation
\begin{equation}
\left(\rho,r \right)  \to \left(z,{\bar r} \right)
\label{}
\end{equation}
of \req{dsdduals} to the Fefferman-Graham form:
\begin{equation}
ds^2 = \frac{\ell_{5}^{2}}{z^2}(dz^2 + \, \Gamma_{\mu\nu}({\bar r}, z)\,dx^\mu\, dx^\nu)
\label{}
\end{equation}
Once this is attained, the series coefficients in the expansion of $\Gamma_{\mu\nu}$ as a function of $z$ as in \req{Gamexp} can be used to determine the stress tensor. In particular, in $d=4$ one has from \cite{deHaro:2000xn}
\begin{eqnarray}
4\pi\,G_N^{(5)} T_{\mu\nu}/\ell_{5}^{3} &=& \, \Gamma_{(4)\mu\nu}  -\frac{1}{8}\,\Gamma_{(0)\mu\nu}\, \left[ \left(\Gamma_{(0)}^{\alpha\beta}\,\Gamma_{(2)\alpha\beta}\right)^2 - \Gamma_{(0)}^{\alpha\beta}\, \Gamma_{(2) \alpha\gamma}\, \Gamma_{(2)\beta\delta}\, \Gamma_{(0)}^{\gamma\delta} \right]
\nonumber \\
&&  \qquad - \frac{1}{2}\, \Gamma_{(2)\mu\alpha}\,\Gamma_{(0)}^{\alpha\beta}\, \Gamma_{(2)\beta\nu} + \frac{1}{4}\, \Gamma_{(2)\mu\nu}\, \Gamma_{(2)\alpha\beta}\, \Gamma_{(0)}^{\alpha\beta}
\label{}
\end{eqnarray}
Explicit calculation for the geometry \req{dsdduals} with $d=4$ gives:
\begin{equation}
 T_\mu^{\, \nu} = c\, H^4 \,\left[\frac{{\cal F}_4(T,H)}{(1-H^2\,r^2)^2}  \; \text{diag}\big\{-3,1,1,1\big\} -\frac{3}{4}\,\text{diag}\big\{1,1,1,1\big\}\right]
\label{}
\end{equation}
As expected the holographic stress tensors provide a smooth interpolation between the theory on de Sitter at temperature $T$ and finite temperature physics in Minkowski spacetime.

\section{Confining Gauge Theory on de Sitter Space}
\label{s:confdS}

Thus far we have concentrated on the behaviour of conformal field theories (with AdS duals) on de Sitter spacetime. However, even working in the static patch at temperatures $T \neq T_{\text dS}$, we have seen that the phase structure of such theories is trivial.  As a result, these states do not quite address our original motivation: investigating phase transitions induced by cosmological evolution.

Thus, we now turn to the task of exhibiting a holographic cosmological phase transition. Ideally one would like to take a phenomenologically relevant theory like QCD and examine its behaviour as one varies the background acceleration. This being a slightly ambitious task, we will instead address a simpler problem  which nevertheless captures much of the basic physics. The idea is to consider a field theory with gravity dual that exhibits a deconfinement transition in Minkowski spacetime, and check whether this theory also exhibits a transition at a critical value of $H$ on de Sitter spacetime. For the theories we consider, the necessary de Sitter analysis was already carried out in \cite{Hutasoit:2009xy} (and earlier work) and indicates that there is indeed a phase transition at a critical value of $H$. Interestingly, we find that this critical value corresponds to a de Sitter temperature $T_{dS}$ {\it lower} than that of the associated phase transition in flat spacetime.

\subsection{Definition of the confining gauge theory}
\label{s:ssconf}

In order to study the phase structure of a confining gauge theory on de Sitter, we first need a field theory with confined and deconfined phases that in addition has a holographic dual. The simplest such theories were described in \cite{Witten:1998qj,Witten:1998zw} and are realized as a Scherk-Schwarz compactification of a CFT on  $\Sp^1_{SS}$. The Scherk-Schwarz boundary conditions are that the fermions in the theory be anti-periodic around the $\Sp^1_{SS}$.  To set the stage for our discussion we will first quickly review the physics of these theories on Minkowski spacetime.  One useful aspect of such Scherk-Schwarz compactifications of CFTs is that one has a geometric picture of confinement: the bulk manifold dual to the theory in the confined phase smoothly caps off in  the bulk effectively producing an IR end to the dual spacetime.\footnote{In the following we will implicitly assume that we are working with a large $N$ gauge theory (or equivalently large central charge $c$ CFT) prior to the Scherk-Schwarz compactification. This in particular, implies that the confined phase will have ${\cal O}(1)$ free energy while the deconfined phase will have ${\cal O}(c)$ free energy, which we take to be the basic order parameter for the confinement/deconfinement transition.}

Generally, a $d+1$-dimensional CFT on ${\bf R}^{d,1}$  has a conformally invariant vacuum state which via the gauge/gravity duality, corresponds to a  dual gravitational background \AdS{d+2}:
\begin{equation}
ds^2 = \left( \frac{u}{\ell_{d+2}} \right)^2 \left(-dt^2 + (dx_i)^2 + (dx_d)^2 \right) + \left( \frac{\ell_{d+2}}{u} \right)^2 \, du^2
\label{adspoin}
\end{equation}
where $i = 1,\cdots, d-1$. Now, we can consider the field theory compactified on $\Sp_{SS}^1$ by taking the $x_d$ direction compact with size $2 \pi \,R$. While one might think that the gravity dual is simply given by periodically identifying $x_d$ in \req{adspoin} with period $2 \pi\,  R$, it turns out the true vacuum state of the field theory is instead given by the AdS-soliton geometry \cite{Horowitz:1998ha} whose metric takes the form:
\begin{equation}
ds^2 = \left( {u \over \ell_{d+2}} \right)^2\, \left( -dt^2 +  (dx_i)^2 + f(u)\,  (dx_d)^2\right)  + \left( {\ell_{d+2} \over u} \right)^2 {du^2 \over f(u)}
\label{confine}
\end{equation}
where
\begin{equation}
f(U) = 1 - {u_0^{d+1} \over u^{d+1}} \; .
\end{equation}
As long as we take
\begin{equation}
u_0 =  \frac{2 \, \ell_{d+2}^2}{ (d+1) \, R}
\end{equation}
the solution smoothly caps off at $u=u_0$. At this point, the circle parametrized by $x_d$ pinches off, so the $x_d$ and $u$ directions form a cigar-type geometry. This pinching off is consistent provided we have anti-periodic boundary conditions for fermions on the circle. As explained in detail in \cite{Witten:1998zw, Horowitz:1998ha}
this solution has lower action when compared to \req{adspoin},  and therefore represents the vacuum state of the field theory. The fact that the solution has an IR end implies that the corresponding field theory is confining (i.e., the gap in the spectrum can indeed be directly related to the capping off of the spacetime).

It is straightforward to show that confining field theories defined in this way undergo a deconfinement transition at a critical temperature
\begin{equation}
T_c = {1 \over 2 \pi \, R} \; .
\end{equation}
The equilibrium state of the field theory at temperature $T$ may be analyzed by considering the Euclidean geometries with boundary behavior as above, but with the Euclidean time compactified on a circle of size $1/T$ with anti-periodic boundary conditions for fermions. One solution with these boundary conditions is the simple analytic continuation of the solution \eqref{conf}, $t \to i \, t_E$ and to take $t_E$ periodic with period $T^{-1}$.

However, since $t_E$ and $x_d$ are now on the same footing (they are compact directions with anti-periodic boundary conditions for fermions) it is clear that we can obtain another solution by the replacements $t_E \leftrightarrow x_d$ and simultaneously exchanging $T^{-1} \leftrightarrow 2 \pi\,  R$. The solution thus obtained has lower action when $T> \frac{1}{2 \pi \,R}$ (i.e. when the $t_E$ circle is smaller than the $x_d$ circle), so we have a phase transition at the critical temperature $T_c$ defined above.

Back in Lorentzian spacetime, the high-temperature solution is
\begin{equation}
ds^2 = \left( \frac{u}{\ell_{d+2}} \right)^2 \left(-f_T(u)dt^2 + (dx_i)^2 +  (dx_d)^2 \right) + \left( {\ell_{d+2} \over u} \right)^2 {du^2 \over f_T(u)}
\label{decon}
\end{equation}
where
\begin{equation}
f_T(u) = 1 - {u_T^{d+1} \over u^{d+1}} \;
\label{deconfT}
\end{equation}
and
\begin{equation}
u_T =  \frac{4 \pi T\,  \ell_{d+2}^2}{d+1} \; .
\end{equation}
The solution \req{decon} is a black hole with planar horizon topology. The horizon is located at $u=u_T$, and corresponds to a deconfined phase of the field theory (e.g. the Polyakov loop has a non-zero expectation value since the temporal circle is contractible in the Euclidean picture \cite{Witten:1998zw}).

\subsection{Confining gauge theories on de Sitter spacetime}
\label{s:confdual}

Having described a class of confining gauge theories that can be studied using AdS/CFT, one would like to investigate the physics of these theories on de Sitter spacetime. Since the field theories were defined as conformal field theories on $\Sp^1_{SS}$, the same field theories on dS spacetime correspond to conformal field theories on $\Sp^1_{SS} \times {\text dS}_d$. Using the basic AdS/CFT dictionary explained at the beginning of \sec{s:cftds}, such theories should be described by the dual gravitational theory on an asymptotically locally AdS spacetime whose boundary geometry is $\Sp^1_{SS} \times {\text dS}_d$. While the circle and the dS spacetime both have a characteristic length scale that can be varied, physical observables only depend on the ratio of these, since the original field theory is conformal. In the language of the confining gauge theory, the physics depends only on how large the Hubble parameter $H$ is compared with the confinement scale $\Lambda_c = \frac{1}{R}$.

In general, as discussed earlier,  for a quantum field theory on de Sitter spacetime, the de Sitter-invariant equilibrium state for a given $H$ can be obtained by analytically continuing the Euclidean vacuum state on the round sphere. For the current case of interest, the relevant Euclidean field theory is then a CFT on $\Sp^d \times \Sp^1_{SS}$.  The AdS/CFT analysis of such a field theory is well known  \cite{Witten:1998zw} and is intimately tied with the observation of Hawking and Page \cite{Hawking:1982dh}. In particular, there are  two asymptotically locally \AdS{} gravity solutions whose boundary geometry is the desired manifold $\Sp^1_{SS}  \times {\text dS}_d$. Moreover, as one might expect these solutions exchange dominance at  a critical value of $R_{\Sp^1_{SS}}/R_{{\text dS}_d}$.

It remains only to write down the Euclidean solutions and to Wick rotate them to Lorentzian solutions with de Sitter boundaries.  This construction is well-described in \cite{Ross:2004cb,Hutasoit:2009xy} and references therein. We will therefore restrict ourselves to a general review these geometries.

\subsubsection{Saddle points of the Euclidean action}
\label{s:euclds}

Parameterizing the $\Sp^1_{SS}$ by a coordinate $\chi$ with period $2 \pi R$, the two (Euclidean) solutions  of interest are given as:
\begin{equation}
ds^2 = {\rho^2 \over \ell_{d+2}^2} \left(f(\rho) \, d\chi^2 + {1 \over H^2} \, d\Omega_{d}^2 \right) + \left( {\ell_{d+2} \over \rho} \right)^2 {d\rho^2 \over f(\rho)}
\label{adssph}
\end{equation}
The function $f(\rho)$ is what determines the nature of the solution and its regime of dominance. The solution which has lower action for a large $\Sp^d$, i.e., for small $H\, R$  is the Euclidean Schwarzschild-\AdS{d+2} black hole for which:
\begin{equation}
f(\rho) = 1 + \frac{H^2 \ell_{d+2}^4}{\rho^2}  - \frac{\rho_+^{d+1}}{\rho^{d+1}}\, \left(1+\frac{H^2 \ell_{d+2}^4}{\rho_+^2}\right)
\label{sol1}
\end{equation}
This solution has topology $\Sp^d \times {\bf R}^2$ for the $\Sp^1$ smoothly contracts to a point as $\rho \to \rho_+$ provided one chooses
\begin{equation}
R =  \frac{2\, \rho_+ \, \ell_{d+2}^2}{(d+1) \, \rho_+^2 + (d-1)\, \ell_{d+2}^4 H^2} \ .
\label{Rcons}
\end{equation}
On the other hand the $\Sp^d$ size remains finite everywhere in the spacetime.  The second solution, is just \AdS{d+2} which has lower action for a small $\Sp^d$ is given by \req{adssph} with:
\begin{equation}
f(\rho) = 1 + \frac{H^2 \ell_{d+2}^4}{\rho^2}
\label{sol2}
\end{equation}
This has topology ${\bf R}^{d+1} \times \Sp^1$ since here the $\Sp^1$ remains finite while the sphere $\Sp^d$ contracts to a point at $\rho=0$.

Note that the solutions \req{sol1}  where the $\Sp^1_{SS}$ contracts smoothly in the interior of the spacetime exist only for a range of the radius of the Scherk-Schwarz circle. In particular, from \req{Rcons} we learn that
\begin{equation}
R_\text{max} = \frac{1}{\sqrt{(d-1)\,(d+1)}}\, \frac{1}{H}
\label{}
\end{equation}
Moreover, the relation between $R$ and $\rho_+$ is double-valued; for a given value of $R$ we have two different choices of $\rho_+$ which translates to two distinct solutions.  On the contrary the second set of solutions where the $\Sp^d$ shrinks to zero size at $\rho =0$ (smoothly) exist for all values of $R$.

The Wick rotation of the Euclidean black holes gives the AdS bubble of nothing solutions of \cite{Aharony:2002cx,Balasubramanian:2002am,Cvetic:2003zy} and the Wick rotation of thermal AdS gives the Ba\~nados black hole \cite{Banados:1997df}.  The Ba\~nados black hole dominates at large $HR$ while the bubble of nothing dominates at small $HR.$  To determine which of the solutions dominates the Euclidean quantum gravity path integral, one needs to compute the on-shell action. Happily this computation is the standard computation carried out originally in $d= 2$ in \cite{Hawking:1982dh} and for general $d$ in \cite{Witten:1998zw}. First of all on can eliminate the class of solutions \req{sol1} with $\rho_+ \le H\, \ell_{d+2}^2$ as they correspond to unstable saddle points of the Euclidean path integral.\footnote{This can be seen by noting that for $\rho_+ \le H\, \ell_{d+2}^2$ that the Euclidean fluctuation operator (Lichnerowicz operator) has a single negative mode in its spectrum \cite{Prestidge:1999uq,Hubeny:2002xn}.} From the  Euclidean action computation of \cite{Witten:1998zw}, one can then learn that the large $\rho_+$ solution \req{sol1} and the second solution \req{sol2} with $\rho_+ =0$ exchange dominance at a critical radius. More specifically, the critical ratio of circle radius to sphere radius $R_{\Sp^1_{SS}}/R_{\Sp^d}$ in the boundary geometry for which the transition occurs is given by
\begin{equation}
R_{\Sp^1_{SS}}/R_{\Sp^d} = {1 \over d} \; .
\end{equation}
In terms of the parameters in the metric above, this translates to
\begin{equation}
H = \frac{1}{d\, R} \; .
\end{equation}
In terms of the de Sitter temperature $T_{dS}$ and the critical temperature $T_c$ for the Minkowski space phase transition, we have
\begin{equation}
T_{dS} = {T_c \over d} \; .
\end{equation}
Perhaps surprisingly, the transition occurs when the de Sitter temperature is lower by a factor of $d$ than the critical temperature in Minkowski space!

\subsubsection{Lorentzian geometries dual to confining theories on de Sitter}
\label{s:confinter}

Having seen the Euclidean description of confining theories on de Sitter, we now return to discussing the Lorentzian geometry and investigate the properties of the two solutions, \req{adssph} with the function $f(\rho)$ being given by \req{sol1} and \req{sol2} respectively. The relevant spacetimes in coordinates covering the static patch of de Sitter are simply
\begin{equation}
ds^2 =  {\rho^2  \over \ell_{d+2}^2}  \left(-(1-H^2\,r^2) \, dt^2 + \frac{dr^2}{1-H^2\, r^2}  +r^2 \, d\Omega_{d-2}^2 +  f(\rho) \, d\chi^2 \right) + \left( {\ell_{d+2} \over \rho} \right)^2 {d\rho^2 \over f(\rho)} \ .
\label{conf}
\end{equation}

As mentioned earlier, for $f(\rho)$ given by \req{sol1} we have the AdS ``bubble of nothing'' spacetime \cite{Aharony:2002cx}, with spacetime ending smoothly at the radius $\rho = \rho_+$ defined in \req{Rcons}. From the point of view of the confining gauge theory, the existence of a bubble of nothing is quite natural: the spacetime has to cap off in the IR, and if the geometry on which the field theory lives is quite flat relative to the QCD scale, one might expect that the full spacetime to look like a ``thickened" version of the boundary. This solution smoothly goes over to the zero-temperature Minkowski solution (\ref{conf}) for $H \to 0$, as we should expect.

On the other hand for $H R > 1/d$, the solution with lower Euclidean action takes the same form as in \req{conf}, but now we take $f(\rho)$ as given by the second solution \req{sol2} above. This solution has a horizon at $\rho=0$, so immediately, we see that it shares one feature with the deconfined solution (\ref{decon}). However, the nature of the horizon is somewhat different, since the entire de Sitter factor in the metric goes to zero at this point. One might worry about the regularity of the spacetime here, but in fact the spacetime is non-singular at this horizon and is in fact just the topological AdS black hole discovered by Ba\~nados \cite{Banados:1997df}. We review some aspects of this in \App{s:banados} for completeness.

\subsection{A phase diagram for confining gauge theories in de Sitter}
\label{s:ordpar}

Our discussion so far considers states defined on global de Sitter, thus imposing the restriction $T = T_\text{dS} = \frac{H}{2\pi}$.  It is interesting to remove this restriction and to consider the theory in the static patch of de Sitter for $T \neq T_\text{dS}$. In particular, we wish to prove that the phase transition described above and studied in \cite{Hutasoit:2009xy} for confining theories on de Sitter is continuously connected to the analogous transition for the same theory on flat space.

With $T\neq  T_{\text dS}$ one can consider more generally the phase diagram for our theory on a static patch of de Sitter spacetime, as a function of $T$ and $H$ (as discussed in \sec{s:staticp}, \sec{s:2dds} and \sec{s:hdcft}). For our confining theory in a regime where the theory has a weakly-curved dual gravity description, the equilibrium phase for a general $T$ and $H$ is related to the minimum-action Euclidean geometry satisfying Einstein's equations with negative cosmological constant and boundary geometry $\Sp^d_\chi \times \Sp^1$ (see \sec{s:staticp} for the definition of $\Sp^d_\chi$). This defines a family of Euclidean geometries depending on two parameters: the angle $\chi$ which encodes the temperature, and the curvature of the $\Sp^d$, which encodes the acceleration parameter.

The topology of the associated Euclidean solution provides an order parameter that divides the $T$-$H$ phase diagram into at least two parts. The region for which the associated Euclidean solution has topology $\Sp^d \times {\bf R}^2$ contains both the low-temperature Minkowski phase and the low-acceleration de Sitter phase, while the topology ${\bf R}^{d+1} \times \Sp^1$ region contains both the high-temperature Minkowski phase and the high-acceleration de Sitter phase. Thus, the former two phases are cleanly distinguished from the latter two phases by an order parameter.

Further, it is clear that the low-acceleration de Sitter phase is smoothly connected to the confined phase of the dual theory on flat space. This follows since in the $H \to 0$ limit the dual gravity solution in the low-acceleration de Sitter phase goes smoothly into the solution \req{confine} dual to the Minkowski-space confined phase. Thus it is certainly correct to identify the $H < H_c$ phase in dS spacetime with the Minkowski space confined phase.

\begin{figure}[t]
\begin{center}
\includegraphics[scale=0.8]{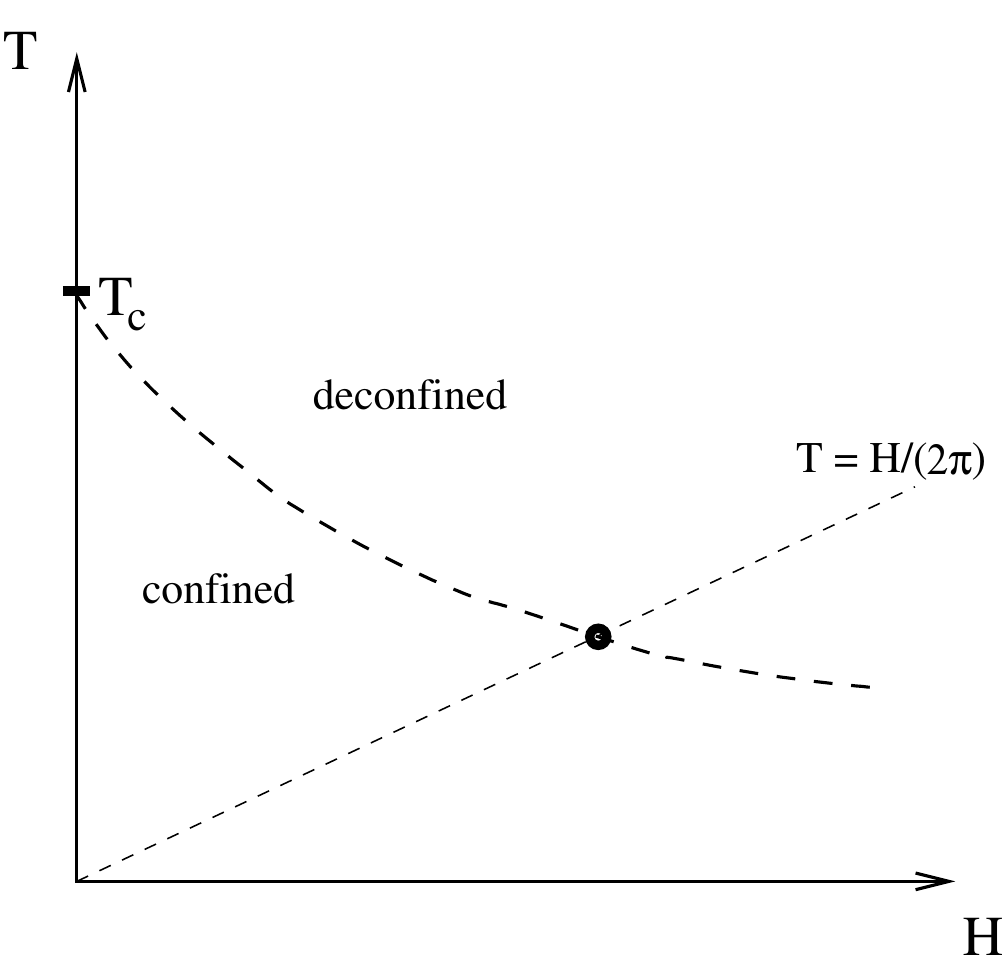}
\caption{Proposed phase diagram for confining gauge theory on de Sitter spacetime. This thin dashed line represents the de Sitter-invariant state with $T=\frac{1}{2\pi}\, H$. The diagram assumes the absence of phase transitions other than the thick dashed line representing a change in topology of the dual Euclidean solution from $\Sp^d \times {\bf R}^2$ to ${\bf R}^{d+1} \times \Sp^1$. Also note that the transition temperature does down as we increase the cosmological acceleration.}
\label{f:confdeconf}
\end{center}
\end{figure}

It remains to show that the Euclidean solution corresponding to the high-acceleration de Sitter phase smoothly connected by a change in parameters to the deconfined phase in Minkowski space. To do this, we would like to find an interpolating solution similar to the one we described in \sec{s:cftds} dual to a conformal field theory at general $T$ and $H$. Note that for the confining theory on \dS{d} $\times \, \Sp^1$, the  bulk solution \req{conf}, \req{sol2} has very similar properties to those associated to  deconfined phases on Minkowski space \cite{Hutasoit:2009xy}, namely the ${\bf R}^{d+1} \times \Sp^1$ topology and the presence of the black hole horizon. It is therefore natural to conjecture that this phase also smoothly carries over as we de-tune the temperature from the value set by the background cosmological constant. In particular, the phase diagram of the system as a function of $T$ and $H$ is expected to be as described  in \fig{f:confdeconf}.

The rest of this section is devoted to providing some modest evidence that a high-temperature interpolating solution exists, and laying out a scheme to verify the conjectured diagram in more detail.  The main idea is simply to expand perturbatively in powers of $H$ around the high temperature solution \req{decon} in Minkowski space  (for simplicity, we focus on the case $d=2$).  This demonstrates that the phase persists for $H> 0$, though numerical work appears to be needed to probe the regime $T \sim H/2\pi$, and thus to fully connect the Ba\~nados black hole with the deconfined phase on flat space.  Given the high symmetry of the desired solutions, one may hope that such numerical investigations will be completed in the near future.

\subsubsection{Duals for confining gauge theory on  de Sitter at $T \neq T_{\text dS}$}
\label{s:pertsolH}
We begin with an ansatz that should be general enough to include the desired solution which interpolates between the Minkowski space deconfined solution \req{confine} and the de Sitter deconfined phase \req{conf}, \req{sol2}.

For simplicity, we restrict  to the case where the field theory lives on \dS{2} $\times \Sp^1$ and furthermore fix the radius of the Scherk-Schwartz circle, $R=1$. We thus seek asymptotically locally \AdS{4} spacetimes which solve Einstein's equations with a negative cosmological constant.  We take as our metric ansatz:
\begin{equation}
ds^2 = u^2 \left\{ - f_t (u,r)\, dt^2 + \frac{dr^2}{ 1 - H^2\, r^2} + e^{j(u,r)} \,d\chi^2 \right\} + \frac{du^2}{u^2 \, f_u(u,r)} \; .
\label{fgjans}
\end{equation}
In order that this agrees with the thermal solution with a Minkowski boundary for $H=0$, we demand that
\begin{eqnarray}
f_t(u,r) &=& 1 - {u_T^3 \over u^3} + {\cal O}(H^2) \cr
f_u(u,r) &=& 1 - {u_T^3 \over u^3} + {\cal O}(H^2) \cr
j(u,r) &=& {\cal O}(H^2)
\end{eqnarray}
On the other hand, in order that the boundary geometry is de Sitter times a circle, we demand that for $u \to \infty$,
\begin{eqnarray}
f_t(u,r) &\to& 1 - H^2 r^2 \cr
f_u(u,r) &\to& 1 \cr
j(u,r) &\to& 0
\end{eqnarray}

It is in fact easy to find a particular class of approximate solutions for the ansatz \req{fgjans} around flat space. To wit,  one solves Einstein's equations to determine $f_t, f_u$, and $j$ perturbatively in $H^2$, but exactly as functions of $u$ and $r$. Up to order $H^2$, we find that a particular solution obeying the conditions above (for simplicity assuming that $j(u,r)= j(u)$ is independent of $r$) is
\begin{eqnarray}
f_t(u,r) &=& 1 - {u_T^3 \over u^3} + H^2 \left\{{C_1 \over u^3} +  \left( 1 + {u_T^3 \over 2 u^3} \right) \left(-r^2 + A(u) + {{\cal C} \over 2} B(u) \right) - {1 \over u^2} + {3 {\cal C} \over 2}  {u_T^3 \over u^3}  \right\} + {\cal O} (H^4)\cr
f_u(u,r) &=& 1 - {u_T^3 \over u^3} + H^2 \left\{{C_1 \over u^3} + {3 u_T^3 \over 2 u^3} \left(-r^2 + A(u) + {{\cal C} \over 2} B(u) \right) + {1 \over u^2}  \right\} + {\cal O} (H^4)\cr
j(u,r) &=& H^2 \left\{ A(u) + {\cal C} B(u) \right\} + {\cal O} (H^4)
\end{eqnarray}
where
\begin{eqnarray}
A(u) &=& {2 \sqrt{3} \over 3 u_T^2} \arctan({\sqrt{3} \over 3} + {2 \sqrt{3} u \over 3 u_T}) -{1 \over 3 u_T^2} \ln \left({(u-u_T)^3 \over u^3 - u_T^3} \right)  - {\sqrt{3} \pi \over 3 u_T^2} \cr
B(u) &=& \ln \left(1 - {u_T^3 \over u^3} \right) \; .
\end{eqnarray}
Here, $C_1$ and ${\cal C}$ are parameters. The parameter $C_1$ can be absorbed into a redefinition of $u_T$ at order $H^2$ (which amounts to a change of coordinates). In fact one can eliminate $C_1$ and set $u_T = 1$ by rescaling coordinates. However, it is easy to check that the solution thus obtained does not have constant bulk surface gravity, implying that there is a potential problem with this geometry being dual to a CFT on \dS{d} in equilibrium at $T\neq T_{\text dS}$.

Given that the particular solution described above is not regular, we would have to do a bit more work to extract the solution which is relevant for the holographic dual. Consider starting with the particular solution found above for $C_1 = {\cal C}=0$
\begin{eqnarray}
f_t(u,r) &=& 1 - {u_T^3 \over u^3} + H^2 \left\{  \left( 1 + {u_T^3 \over 2 u^3} \right) \left(-r^2 + A(u)  \right) - {1 \over u^2}  + F(u,r) \right\} + {\cal O} (H^4)\cr
f_u(u,r) &=& 1 - {u_T^3 \over u^3} + H^2 \left\{ {3 u_T^3 \over 2 u^3} \left(-r^2 + A(u) \right) + {1 \over u^2} + G(u,r) \right\} + {\cal O} (H^4)\cr
j(u,r) &=& H^2 \left\{ A(u)  + J(u,r) \right\} + {\cal O} (H^4) \ .
\end{eqnarray}
We would like to find the most general correction terms $F,G,J$ such that our conditions above are satisfied and Einstein's equations are satisfied to order $H^2$. The resulting equations for $F$, $G$, and $U$ are homogeneous linear differential equations, so the general solution may be written as a linear combination of some complete set of solutions. We first note that starting with the $(r,u)$ component of Einstein's equations gives an equation that we can integrate with respect to $r$, and the resulting equation can be used to solve for $G(u,r)$ in terms of $F$ and $J$,
\begin{equation}
G = -{1 \over u^2 (4u^3 - u_T^3)} \left\{2(u^3 - u_T^3)^2 J_u + 2u^3(u^3 - u_T^3) F_u - 3 u^2 u_T^3 F \right\}
\end{equation}
After this, we find that the $(r,r)$ component is automatically satisfied. Of the remaining three equations coming from the $(u,u)$ component, the $(t,t)$ component, and the $(\chi,\chi)$ component, it is straightforward to check explicitly that only two are independent. The $(u,u)$ equation gives:
\begin{eqnarray}
\label{first}
 F_{rr} +   f(u) J_{rr} &=& A(u) F_u + B(u) F + C(u)  J_u
\end{eqnarray}
where $f(u) = 1 - u_T^3/u^3$ and
\begin{equation}
\left( \begin{array}{c}
A(u) \cr B(u) \cr C(u)
\end{array} \right)  =
{u^3 - u_T^3  \over u^3 (4u^3 - u_T^3)}
\left(\begin{array}{c}
2u^3 (2 u^3 + u_T^3)
\cr 6 u^2 u_T^2
\cr {1 \over 2} (8 u^6 - 16 u^3 u_T^3 - u_T^6)
\end{array} \right)
\end{equation}
The  $(\chi,\chi$ equation contains terms with $F_{rru}$ and $J_{rru}$, but these appear in the combination $ J_{rru} f(u) + F_{rru}$ so we can use the equation (\ref{first}) to eliminate them. The resulting equation is
\begin{equation}
\label{second}
F_{rr} =  a(u)F_{uu} + b(u) F_u + c(u)F + d(u)J_{uu} + e(u) J_u
\end{equation}
where
\begin{equation}
\left( \begin{array}{c}
a(u) \cr b(u) \cr c(u) \cr d(u) \cr e(u)
\end{array} \right)  =
{u^3 - u_T^3 \over 2u^3(4u^3 - u_T^3)^2}
\left(\begin{array}{c}
 -4u^4(u^3 - u_T^3)(4u^3-u_T^3)
\cr -24u^3(u^3 - u_T^3) (2u^3 - u_T^3)
\cr 24 u^2 u_T^3 (u^3 - u_T^3)
\cr 2 u (u^3 - u_T^3)(2u^3 + u_T^3) (4u^3 - u_T^3)
\cr 80u^9 - 96u^6 u_T^3 + 42 u^3 u_T^6 + u_T^9
\end{array} \right)
\end{equation}
We would now like to derive a basis of solutions for these equations. First, we consider a formal solution of the form
\begin{eqnarray}
J(r,u) &=& j(r)J_0(u) + j''(r)J_1(u) + j^{(4)}(r)J_2(u)+ \dots \cr
F(r,u) &=& j(r)F_0(u) + j''(r)F_1(u) + j^{(4)}(r)J_2(u)+ \dots \; ,
\label{formal}
\end{eqnarray}
defined in terms of an arbitrary function $j(r)$. Substituting these expressions, we obtain the simple recursion relations
\begin{eqnarray}
 A(u) (F_n)_u + B(u) F_n + C(u)  (J_n)_u &=& F_{n-1} +   f(u) J_{n-1} \cr
a(u)(F_n)_{uu} + b(u) (F_n)_u + c(u)(F_n) + d(u)(J_n)_{uu} + e(u) (J_n)_u  &=& F_{n-1}
\end{eqnarray}
where the right side vanishes for $n=0$. These may be solved recursively by using the first equation to solve algebraically for $(J_n)_u$ at each order $n$ and then substituting into the second equation to get a differential equation for $F_n$. The $n=0$ equation is a homogeneous differential equation,
\begin{equation}
 \alpha(u) (F_0)_{uu} + %
\beta(u)  (F_0)_{u} + \gamma(u) F_0 = 0
\end{equation}
where
\begin{eqnarray}
\alpha(u) &=& u^2(u^3 - u_T^3)(8u^6 - 16 u^3 u_T^3 - u_T^6) \cr
\beta(u) &=& u(32 u^9 - 120 u^6 u_T^3 + 54 u^3 u_T^6 + 7 u_T^9) \cr
\gamma(u) &=& - 9 (4 u^3 - u_T^3) u_T^6 \; .
\end{eqnarray}
There is one solution of this that goes to zero for $u \to \infty$, and this gives
\begin{eqnarray}
F_0 &=& (1 + {u_T^3 \over 2 u^3})\ln(1-{u_T^3 \over u^3}) + 3 {u_T^3 \over u^3} \cr
J_0 &=& 2 \ln(1-{u_T^3 \over u^3})
\end{eqnarray}
The differential equation for each remaining $n$ takes the form
\begin{equation}
\alpha(u) F_{uu} + %
\beta(u)  F_{u} + \gamma(u) F(u) = Q(u)
\end{equation}
whose solution can be given explicitly in terms of an integral involving $Q(u)$.

Once we have determined the functions $J_n(u)$ and $F_n(u)$, the expressions (\ref{formal}) provide a formal solution for any choice of $j(r)$. The series will be finite if and only if $j(r)$ is a polynomial. To get a basis of legitimate solutions (even in $r$), we can take $j(r) = 1, \; r^2, \; r^4,\dots$. For example, the first two solutions in this set are:
\begin{equation}
\begin{array}{ll}
J^{(0)}(r,u) = J_0(u) & F^{(0)}(r,u) = F_0(u) \cr
J^{(1)}(r,u) = J_0(u) r^2 + 2 J_2(u) & F^{(1)}(r,u) = F_0(u)r^2 + 2 F_2(u)\cr
J^{(2)}(r,u) = J_0(u) r^4 + 12 r^2 J_2(u) + 24 J_0(u) & F^{(2)}(r,u) = F_0(u) r^4 + 12 r^2 F_2(u) + 24 F_0(u)\cr
\end{array}
\label{expl}
\end{equation}

A nice trick to solve the recursion relations above is to make use of a generating function. If we define
\begin{equation}
{\cal F}(u,x) = \sum_{n=0}^{\infty} x^n F_n(u) \qquad \qquad {\cal J}(u,x) = \sum_{n=0}^{\infty} x^n J_n(u)
\end{equation}
and define $ F_{-1}(u) = J_{-1}(u) = 0$ then the entire set of recursion relations reduces to the pair of ordinary differential equations
\begin{eqnarray}
x {\cal F} +   x f(u) {\cal J}  &=&  A(u) {\cal F}'  + B(u){\cal F}  + C(u)  {\cal J}'    \cr
x {\cal F} &=&   a(u){\cal F}''  + b(u){\cal F}' + c(u){\cal F} + d(u){\cal J}'' + e(u) {\cal J}'
\end{eqnarray}
where prime indicates a derivative with respect to $u$ and $x$ appears only as a parameter.

In the general solution above, we have an infinite set of parameters that are so far unfixed. To understand how these should be fixed, it is helpful to consider an alternate approach, working to all orders in $H$, but perturbatively in ${1 \over u}$, again demanding the $H=0$ behavior and the large $u$ behavior above. In this case, we obtain up to order $u^{-5}$
\begin{eqnarray}
j(u,r) &=& {1 \over u^2} H^2  + {1 \over u^3} H^2 C(r) + {1 \over u^4} (-{H^4 \over 2}) \cr
f_t(u,r) &=& (1 - H^2 r^2) + {1 \over u^3}\left( - {u_T^3 \over \sqrt{1-H^2 r^2}}  - {1 \over \sqrt{1 - H^2 r^2}} \int d \tilde{r}(1 - H^2 \tilde{r}^2)^{3 \over 2} C'(\tilde{r})  \right) \cr
f_u(u,r) &=& 1 +{1 \over u^2} H^2 + {1 \over u^3}\left( - {u_T^3 \over (1-H^2 r^2)^{3 \over 2}}  + C(r) - {1 \over (1 - H^2 r^2)^{3 \over 2}} \int d \tilde{r}(1 - H^2 \tilde{r}^2)^{3 \over 2} C'(\tilde{r})  \right) \nonumber
\end{eqnarray}
We see that the most general solution may be expressed in terms of a single arbitrary function $C(r)$.

The function $C(r)$ should be fixed by demanding that the Euclidean solution has no conical singularity along the curve $f_t(u,r) = 0$. This is a whole function worth of conditions (for each r, there will be some $u_\star(r)$ where $f_t(u_\star(r),r) = 0$, and we could have a potential conical singularity. Demanding the absence of such a singularity typically requires the freedom of a whole function and should thus fix the solution of interest uniquely. In particular, this condition will determine which linear combination of the solution basis above corresponds to the physical solution of interest.

\section{Discussion}
\label{s:discuss}

The main focus of the current work has been to understand the behaviour of strongly coupled field theories in the static patch of de Sitter spacetime for $T \neq T_{dS}$. In particular, we investigated the dynamics of CFTs and confining gauge theories using the gauge/gravity duality.

For conformal field theories on global de Sitter one has a unique vacuum state obtained from the Euclidean Bunch-Davies vacuum.  This vacuum is thermal in nature owing to the fact that a static observer in de Sitter experiences thermal radiation emanating from his cosmological horizon. However, as we have argued it is possible to consider these theories in the static patch of de Sitter at temperatures different from $ T_{\text dS}$, at the expense of introducing a source on or just behind the cosmological horizon.  We have constructed gravity duals for such theories and shown explicitly that the corresponding solutions in the bulk spacetime are manifestly regular. In particular, an interesting feature of these solutions is the fact that the bulk horizon only knows about the temperature the field theory is at, rather than the de Sitter temperature. This is despite the fact that the bulk horizon remains continuously connected to the de Sitter horizon on the boundary, see \fig{f:ds2btz}.

In addition to the behaviour of CFTs on de Sitter spacetime, we have also investigated the phase structure of confining gauge theories on these geometries. The class of confining theories we have investigated corresponds to those obtained via Scherk-Schwarz compactification of a higher dimensional CFT. The confinement scale in these theories is set at the Kaluza-Klein scale of the Scherk-Schwarz circle. As described in \cite{Hutasoit:2009xy}, for such theories a suitable analytic continuation of the Hawking-Page phase transition, known to occur for thermal CFTs on the Einstein Static Universe, can be used to demonstrate the existence of a  phase transition. This phase transition occurs  for the confining gauge theory  as the acceleration parameter is varied past a critical value.  We provided some modest evidence that this phase transition can be smoothly connected to the confinement/deconfinement transition in flat space by decoupling the temperature $T$ from the Hubble constant $H$.

In realistic cosmological spacetimes, the Hubble parameter changes dynamically, so we expect such phase transitions to occur dynamically in that context. In principle, this process could be studied using AdS/CFT by generalizing the de Sitter boundary metric we have considered here to an FRW-type metric with varying Hubble parameter. The reduced symmetry in that situation would presumably necessitate a numerical study. Alternatively, one could study the dynamics of transitions by working at fixed $H$ but starting in the phase corresponding to larger Euclidean action. In this case, we should expect tunneling from the false vacuum to the true vacuum. Precisely such a process was studied in detail in \cite{Balasubramanian:2005bg}, in the context of decay of $AdS$ space via a bubble of nothing. Reinterpreted in terms of the dual confining gauge theory, the process studied in \cite{Balasubramanian:2005bg} represents a dynamical transition to a confined phase which occurs uniformly in space. More generally, one might expect the transition to proceed less uniformly (e.g. by formation and percolation of bubbles), so it is possible that the gravitational instability considered in \cite{Balasubramanian:2005bg} could be dominated by some less-symmetric process.

\subsection*{Acknowledgements}
\label{s:acks}

We would like to thank Veronika Hubeny for collaboration during the initial stages of the project and for useful discussion. We would also like to thank Vijay Balasubramanian, Simon Ross and Moshe Rozali for useful discussions. MR would like to thank the Amsterdam String Workshop for their hospitality during the course of this project. MR is supported in part by a STFC Rolling Grant and a Theory Special Programme Grant. DM was supported in part by the US National Science Foundation under grants  PHY05-55669 and PHY08-55415 and by funds from the University of California. MVR was supported in part by the Natural Sciences and Engineering Research Council of Canada and the Canada Research Chairs Programme.

\appendix

\section{Geometries dual to static patch of $dS_2$ at arbitrary temperature}
\label{s:ds2sol}

In this appendix we give some basic details regarding the construction of the three dimensional geometries solving the equations of motion derived from \req{bulkact} whose boundary is \dS{2} with a temperature $T$ generically different from $T_\text{dS}$.

Since the boundary metric is static, we expect that the corresponding bulk solutions should also be static. This motivates an ansatz of the form \req{ds2HT}. One can in fact easily show that for such an ansatz, all solutions are required to be of the form:
\begin{eqnarray}
j(r,z) &=& \frac{1}{4} \, (A(r) + B(r)\, z^2)^2 \nonumber \\
f(r,z) &=& \frac{1}{4} \, (C(r) + D(r)\, z^2)^2
\end{eqnarray}
where the functions $A,B,C,D$ depend solely on $r$ and  are subject to additional coupled ordinary differential equations
\begin{eqnarray}
\frac{C'(r)}{A(r)} &=& \frac{D'(r)}{B(r)} \nonumber \\
A(r)\, D(r)+B(r) \,C(r) &=& 2 \,\frac{d}{dr} \left( \frac{C'(r)}{A(r)} \right)
\end{eqnarray}
The solutions corresponding to the CFT on \dS{2} with temperature equal to the de Sitter temperature has
\begin{eqnarray}
A(r) + B(r) \,z^2 &=& \frac{1}{\sqrt{1 - H^2 \,r^2}} - z^2 \,\frac{H^2}{\sqrt{1- H^2\, r^2}} \cr
C(r) + D(r) \, z^2 &=& \sqrt{1 - H^2 \,r^2} - z^2\, H^2\, \sqrt{1 - H^2 \,r^2}
\end{eqnarray}
while the solution corresponding to $H=0$ at temperature $T$ has
\begin{eqnarray}
A(r) + B(r)\, z^2 &=& 1 + (2 \pi \,T)^2 \,z^2 \cr
C(r) + D(r) \,z^2 &=& 1 - (2 \pi \,T)^2 \,z^2 \; .
\end{eqnarray}
We would like to find an interpolating solution corresponding to nonzero $H$ with $T$ not necessarily equal to $T_\text{dS}$. In order that the boundary metric is de Sitter, we must have
\begin{equation}
A(r) = \frac{1}{\sqrt{1 - H^2\, r^2}} \qquad \qquad C(r)  = \sqrt{1 - H^2 \,r^2} \; .
\end{equation}
The temperature is encoded in the periodicity of the Euclidean time, which is taken to be $\beta = T^{-1} =\frac{2 \pi}{u_0}$.

With this choice, we find that the most general solution to the differential equations for $B(r)$ and $D(r)$ is in fact given by
\begin{eqnarray}
B(r) &=& -  \frac{H^2}{\sqrt{1- H^2 \,r^2}} + \frac{\alpha}{(1- H^2 \,r^2)^{3 \over 2}} \cr
D(r) &=& -  H^2 \sqrt{1 - H^2\, r^2} - \frac{\alpha}{\sqrt{1- H^2 \,r^2}}
\end{eqnarray}
where $\alpha$ is an arbitrary constant of integration. This result matches both solutions above in the appropriate limits $H=0$ and $H = 2 \pi \,T $ provided we make the choice
\begin{equation}
\alpha = (2 \pi\, T)^2 - H^2 \; .
\end{equation}
With this choice we  arrive at the final interpolating solution as reported in \req{dsfinal}.

\section{The topological AdS black hole}
\label{s:banados}

Consider the Lorentzian spacetime \req{conf} which is the dual of the confining gauge theory on de Sitter for $H R > \frac{1}{d}$. First of all it is easy to see that the spacetime is regular in the vicinity of $\rho =0$. In fact, as $\rho \to 0$ one encounters the origin of global \AdS{d+2} (in the Euclidean solution).  For the Lorentzian solution a change of coordinates
\begin{equation}
{\hat \rho}= \rho\,  \cosh t \, \qquad  \tau = \rho\,  \sinh t
\label{}
\end{equation}
brings \req{conf}, \req{sol2} to the manifestly regular form:
\begin{equation}
ds^2 = d\chi^2 + d{\hat \rho}^2 - d\tau^2 + {\hat \rho}^2\, d\Omega^2.
\label{}
\end{equation}

 In fact, the metric we have written covers a patch of the ``topological black hole'' solution described by Ba\~nados \cite{Banados:1997df}.  To understand this solution, we note that the global \AdS{d+2} spacetime has a boost symmetry generated by a vector $\xi^a$ (which is one of the generators of $SO(d+1,2)$) that one can use to quotient the geometry. This orbifold construction is similar in spirit to the construction of the BTZ black hole in three dimensions. In fact, one can write the metric \req{adssph}, \req{sol2} in manifestly BTZ form by the following coordinate transformation:
\begin{equation}
\rho^2 = \left(\frac{2\pi\,\ell_{d+2}}{R}\right)^2\, \left( \varrho^2 -  \varrho_+^2 \right) \ , \qquad  \phi = \frac{2\pi}{R} \, \chi \ , \qquad  \varrho_+ = \frac{1}{2\pi}\,\ell_{d+2}\, H\, R
\label{}
\end{equation}
which leads to the metric
\begin{equation}
ds^2 = \frac{\ell_{d+2}^2}{ H^2\,\varrho_+^2}\, ( \varrho^2 - \varrho_+^2)\, \left(-(1-H^2\,r^2) \, dt^2 + \frac{dr^2}{1-H^2\, r^2}  +r^2 \, d\Omega_{d-2}^2  \right) + \frac{\ell_{d+2}^2 \, d \varrho^2}{ \varrho^2-  \varrho_+^2} +  \varrho^2 \, d\phi^2
\label{}
\end{equation}
The only difference from the conventional BTZ metric in 3 dimensions is the fact that the temporal direction has been replaced by a copy of \dS{d}.

To understand the causal structure, it is useful to pass over to Kruskal type coordinates described in \cite{Banados:1997df}. Continuing with the BTZ analogy it is possible to find a  coordinate chart that cover the fully-extended spacetime (essentially by invoking the embedding coordinates for \AdS{d+2} and finding an appropriate representative for the boost generator employed in the quotient). In such a suitable set of  coordinates the metric for \AdS{d+2} takes the form:
\begin{equation}
ds^2 = \ell_{d+2}^2 \left(1 + {r \over r_+} \right)^2 \eta_{\mu \nu} dy^\mu dy^\nu + r^2 d \phi^2
\label{topbh}
\end{equation}
where
\begin{equation}
r = r_+ {1 + y^2 \over 1 - y^2} \qquad \qquad y^2 = \eta_{\mu \nu} y^\mu y^\nu \ ,
\end{equation}
The boost generator in question is then simply $\xi^a = \partial_\phi$ and quotient construction  just involves identifying the coordinate $\phi$ to have  period $2 \pi$. In these coordinates, the boundary of the spacetime is at $y^2 = 1$, the horizon is the light cone $y^2=0$, and we have a ``singularity'' at $y^2 = -1$ where the $\phi$ circle pinches off. The Penrose diagram  is shown in \fig{f:bbh}(b). An interesting feature of this geometry is the absence of a bifurcation surface on the horizon. Since the horizon located at $r = r_+$ (equivalently $y=0$) is simply a copy of \dS{d} we see that one has instead a bifurcation point.

\providecommand{\href}[2]{#2}\begingroup\raggedright\endgroup

\end{document}